\documentclass[letterpaper]{article}

\usepackage[T1]{fontenc}
\usepackage{comment}

\usepackage{geometry}
\usepackage{setspace}
\usepackage{pdfpages}

\usepackage[style=chem-acs, backend=biber]{biblatex}
\usepackage{xpatch}

\xpatchbibdriver{article}
  {\usebibmacro{journal+issuetitle}}
  {\printfield{title}\newunit\newblock
   \usebibmacro{journal+issuetitle}}
  {}{}
\usepackage{graphicx}
\usepackage{float}
\newfloat{scheme}{htbp}{los}
\floatname{scheme}{Scheme}
\floatname{chart}{Chart}
\newfloat{graph}{htbp}{loh}

\usepackage[version = 4]{mhchem} 

\usepackage{placeins}
\usepackage{xcolor}
\usepackage{graphicx}
\usepackage{dcolumn}
\usepackage{bm}
\usepackage{booktabs}
\usepackage{hyperref}
\usepackage{amssymb}

\usepackage{authblk}
\author{S. K. Gregg*} 
\author{D. G. Green$^\dagger$} 

\affil{
 School of Mathematics and Physics, Queen's University Belfast, Belfast BT7 1NN, Northern Ireland, United Kingdom.
}

\date{Email: *s.gregg@qub.ac.uk, $^\dagger$d.green@qub.ac.uk}

\title{Many-body theory predictions of positron binding energies in five-membered heterocycles involving N, O, S and NH substituents.}
\begin{document}

\maketitle

\begin{abstract}
  Positron binding energies and Dyson orbitals for five-membered heterocycles with N, O, S and NH substituents are predicted \emph{ab initio} via many-body theory.
  The positron-molecule correlation potential (self energy) is calculated via solution of Bethe-Salpeter equations that describe the positron-induced polarization of the target and screening of the electron-positron Coulomb interaction at the $GW$@BSE level, the infinite electron-positron ladder series that describes the crucially important process of virtual positronium formation, and the analogous positron-hole ladder series. The all-order calculations employ Gaussian-orbital bases and are implemented in the {\tt EXCITON+} code. 
   The effect of substituting combinations of N, O and S atoms, and the NH group in the molecule's ring is studied, 
   and the role of individual molecular orbitals, many of which are found to significantly contribute to the correlation potential, quantified. 
Analysis of the positron bound-state Dyson orbitals shows that the positron is typically localized next to one or two of the substituents in the ring, with the order of preference N, S, O, then NH, and is also influenced by aromaticity and the presence of double ($\pi$) bonds in the ring. 
\end{abstract}





\section{\label{sec:introduction} Introduction}
One of the early remarkable consequences of the development of quantum mechanics was Dirac’s prediction of the positron and antimatter, now nearly a century ago itself. 
Nowadays, positrons have important use as ultrasensitive probes of defects, surfaces, and porosity in materials \cite{Tuomisto2013,Hugenschmidt2016}; in medical imaging via positron emission tomography including  emerging positronium-based techniques \cite{BassRMP,Moskal2024,JPET2025}; and in astrophysics \cite{Prantzos2011,Fuller:2019,Flambaum:2021,posfun}. They are routinely slowed, trapped, accumulated, and delivered in energy-resolvable beams from traps \cite{RevModPhys.82.2557, Danielson:2015} or nuclear reactor sources \cites{Hugenschmidt2016} for fundamental atomic and molecular physics studies (see e.g., \cite{RevModPhys.82.2557, Green2014,Green2015,  Hofierka2022,Rawlins2023} and references therein), and are also used to form more complicated antimatter, namely positronium (Ps)  and antihydrogen \cite{cassidy2018, alpha1,alpha2,ATRAP:2012,Amole:2012,Amole:2014,GBAR,Ahmadi:2016,Ahmadi:2017,Malbrunot18,Baker2021,Amsler21,Moskal2021,Adrich2023,Anderson2023}, which are under ongoing intensive investigation to study fundamental symmetries and gravity. 

Positrons also hold promise as a unique and novel form of molecular spectroscopy.  
It is now well established that positrons can bind to molecules via capture into vibrational Feshbach resonances, in which the positron binds whilst simultaneously vibrationally exciting the molecule, leading to dramatic, experimentally resolvable enhancements in annihilation rates \cite{RevModPhys.82.2557}.
The resulting energy-resolved annihilation spectra are molecule-specific and probe vibrational dynamics beyond those accessible to conventional optical spectroscopy, including non-dipole, multimode coupling and intravibrational molecular redistribution \cite{RevModPhys.82.2557,Gribakin2017}. 
Positron binding and annihilation resonances can
also probe the internal molecular vibrational temperature \cite{Swann:2020}.
The \emph{ab initio} description of the vibrational-Feshbach resonance annihilation spectra is a formidable open problem, requiring accurate description of the positron scattering state, and the coupling of the electronic and vibrational degrees of freedom. 
However, the theoretical description of the positron-molecule bound state even in the absence of vibrations is challenging, owing to the strong many-body correlations that characterize the positronic/electronic part of the problem. They include positron-induced polarization of the molecular electron cloud, screening of the electron-positron Coulomb interaction, and the crucial non-perturbative process of virtual positronium formation \cite{PhysScripta.46.248,PhysRevA.52.4541,Gribakin:2004,Hofierka2022}. These effects have an overwhelming effect, acting against the positron-nuclear repulsion to significantly modify scattering properties, strongly enhance annihilation, and enable and enhance binding and positronic bonding \cite{Cassidy2024_2}. The problem is substantially more challenging than its all-electron counterpart. 

For a molecule with dipole moment larger than 1.625 D, positron binding occurs even at the static level of theory \cite{Crawford1967}, and the correlations act to enhance the binding (as opposed to \emph{enabling} it, in the case of weakly and non-polar molecules). 
Many calculations therefore predominantly focused on strongly polar molecules.
Theoretical approaches included the
configuration interaction \cite{Buenker2008,Gianturco2006}, any-particle molecular-orbital \cite{Charry2014,Romero2014} and multi-component molecular orbital methods \cite{Ozaki2021,Koyanagi2012,Tachikawa2011,Tachikawa2014}. However, where measurements were available, these achieved agreement to within 25\% of experiment at best.
Estimates of positron-molecule binding energies have also been produced through positron scattering calculations with the Schwinger multichannel method \cite{Barbosa2019,Moreira2019,Moreira2024,Frighetto2025} and R-matrix method \cite{Danby1988,Edwards2021,Graves2022}. 
Highly accurate calculations using the explicitly correlated Gaussian \cite{Bubin2004} and quantum Monte Carlo \cite{Kita2009,Yamada2014} methods have been performed, but are limited to small molecules due to their high computational expense. 
The isotropic model potential approach of Swann and Gribakin provided positron binding energies in good agreement with experiment for alkanes and chlorinated hydrocarbons \cite{Swann2019,Swann2020,Swann2021}, as did that of Tachikawa et al., which also accounted for the the potential ansiotropy \cite{Tachikawa:MCP_a,Ashiba2025}, but relies on the choice of some free parameters for each species of atom in the system, heavily restricting its predictive capability. 
Early analysis of experimental data suggested an empirical relation between the positron-molecule binding energy and global molecular properties, namely the dipole moment, isotropic polarizability, ionization energy and number of $\pi$ bonds \cite{Danielson09}. 

In 2022, we developed the \emph{ab initio} diagrammatic many-body theory of positron-molecule binding, achieving the first quantitative agreement with measured positron binding energies and providing fundamental insight that highlighted the need for \emph{ab initio} calculations that properly accounted for the correlations but also the anisotropy of the interactions and molecular symmetry. 
The approach combines Gaussian-orbital electronic-structure methods with an all-order treatment of electron-positron correlations via Bethe-Salpeter equations, building on the early foundations of atomic many-body theory laid in the 1960s by Kelly \cite{Kelly1963,Kelly1964,Kelly1968} and later extended to positron-atom systems by Amusia \cite{Amusia:Pos:MBT:He}, and Flambaum, Dzuba, Gribakin and colleagues (see e.g., \cite{PhysScripta.46.248,PhysRevA.52.4541,Gribakin2004,harabati2014identification}), and through related diagrammatic approaches in quantum chemistry by Cederbaum \cite{Cederbaum-elecpos}. 
Our approach has been shown to provide good to excellent agreement with experimental positron binding energies for polyatomic halogenated, ringed, and other molecules \cite{Hofierka2022,Cassidy2024,Hofierka2024,ArthurBaidoo2024,Gregg2025}, and has been extended to describe positron scattering and annihilation in atoms and molecules \cite{Rawlins2023,Hofierka2023,Gregg2025Gamma} and to predict new types of positronically-bonded molecules \cite{Cassidy2024_2}. 
After, a neural network variational Monte Carlo approach was used successfully to describe positron binding in benzene and other small molecules \cite{Cassella2024}. Recent calculations were also performed using the configuration interaction and diffusion Monte Carlo methods for atoms and small molecules \cite{Upadhyay2024}. Moreover, the very recently developed positron coupled-cluster (including by one of us) \cite{Rosario2026} (POS-CCSD; including singles and double electronic excitations, and simultaneous single positron excitations), holds promise as a powerful accurate \emph{ab initio} method, but owing to the computational expense, has not yet achieved converged results for the polyatomic molecules it considered. 

In this work
we apply our state-of-the-art diagrammatic many-body theory framework to predict positron binding to a series of 13 five-membered heterocycles containing N, O, S, and NH substituents. These are true predictions, with no prior experimental or comparable theoretical data available.
Four of the molecules have a single heteroatom in the ring: tetrahydrofuran (\ce{C4H8O}), thiophene (\ce{C4H4S}), tetrahydrothiophene (\ce{C4H8S}) and pyrrolidine (\ce{C4H8NH}); three contain multiple nitrogen heteroatoms: imidazole (\ce{C3H4N2}), pyrazole (\ce{C3H4N2}) and triazole (\ce{C2H3N3}); and the remaining six molecules each have two distinct substituted atoms in the ring: oxazole (\ce{C3H3NO}), isoxazole (\ce{C3H3NO}), thiazole (\ce{C3H3NS}), isothiazole (\ce{C3H3NS}), 1,2-oxathiole (\ce{C3H4OS}) and 1,3-oxathiole (\ce{C3H4OS}). 
By systematically varying heteroatom identity and placement, we elucidate how molecular composition, polarity, and $\pi$-electron structure govern positron binding energies and bound-state wavefunctions, and we analyze the orbital-resolved contributions to the positron-molecule correlation potential.
The heterocyclic motifs considered here are ubiquitous in chemically and biologically relevant systems: oxazole, thiazole, and triazole rings appear in PET radiotracers, while pyrrole and pyrrolidine units occur in amino acids and other biomolecules. Beyond their intrinsic importance, the present results substantially expand the range of systems accessible to modern \emph{ab initio} many-body theory and provide insight and benchmarks for alternative computational approaches to the positron-molecule and many-fermion problem, including emerging data-driven methods which for the positron are challenging if not unfeasible due to the current small amount of positron-molecule binding data. 

\section{Many-body theory and its implementation for heterocycles}\label{sec:methodology}
\subsection{Many-body theory of positron-molecule binding}

The many-body theory approach employed here has been detailed extensively previously, see e.g.,  \cite{Hofierka2022, Cassidy2024, ArthurBaidoo2024}.
Briefly, we calculate the positron binding energy $\varepsilon_{\rm b}$ and bound-state wavefunction by solving the Dyson equation \cite{mbtexposed,fetter},
\begin{equation}\label{dyson_equation}
   \left( \hat{H}_0 + \hat{\Sigma}_{\varepsilon}\right) \psi_{\varepsilon}(\mathbf{r}) = \varepsilon \psi_{\varepsilon}(\mathbf{r}),
\end{equation}
where $\psi_{\varepsilon}(\mathbf{r})$ is the quasiparticle positron wavefunction with energy $\varepsilon$, $\hat{H}_0$ is the Hamiltonian of the positron in the static, Hartree-Fock (HF) field of the molecule, and $\hat{\Sigma}_{\varepsilon}$ is the positron-molecule correlation potential (self-energy operator), which accounts for positron-molecule correlations. It is non-local, acting as $(\hat{\Sigma}_{\varepsilon}\psi_{\varepsilon})({\bf r}) = \int \Sigma_{\varepsilon}({\bf r},{\bf r}')\psi_{\varepsilon}({\bf r}') d{\bf r}'$ and is energy dependent, meaning that calculations of binding energies require the self-consistent solution of the Dyson equation.
The self-energy is calculated as a diagrammatic (`Dyson') expansion in residual electron-electron and electron-positron interactions. 
The intractability of the $N$-electron plus positron Schr\"odinger equation translates in the many-body approach to the inability to calculate the full infinite diagrammatic series. However, the dominant diagrams, including infinite classes of diagrams, can be identified and resummed. We include three such infinite classes in our calculation of $\Sigma$ that account for the dominant physics. 
The first is the so-called $GW$@BSE self-energy $\Sigma^{GW}$, which describes the positron-induced polarization of the electron cloud and screening of the positron-electron Coulomb interaction and electron-hole attraction via the infinite random phase with exchange approximation series, but also including screened Coulomb interactions in the electron-hole bubble diagrams. Note this series includes time-forward and time-reversed bubbles and exchange diagrams in which multiple electron-hole pairs can be present simultaneously, and takes account of important electron-electron correlation. 
However, unlike the all-electron problem, for the positron-molecule problem the $GW$@BSE contribution alone is known to be severely deficient. 
Of crucial importance in the positron-atom or molecule problem is the additional process of virtual positronium formation (in which a molecular electron temporarily tunnels to the positron) leading to a strong short-range attractive interaction \cite{PhysScripta.46.248,PhysRevA.52.4541,Gribakin2004,Green2013,Green2014,Green2015,Hofierka2022,Hofierka2024,Cassidy2024,ArthurBaidoo2024}. It is accounted in $\Sigma$ via the infinite ladder series of electron-positron interactions $\Sigma^{\Gamma}$; its importance relative to the all-electron problem lies in the fact that successive rings of the ladder in the positron-molecule problem contribute with the same sign, whereas in the all-electron problem they alternate leading to significant cancellation. In practice we calculate the ladder series using both bare and screened electron-positron Coulomb interactions. The third contribution to $\Sigma$ that we account for is the analogous positron-hole infinite ladder series $\Sigma^{\Lambda}$, which describes the positron-hole repulsion and which in practice slightly mitigates the virtual-Ps attraction.
Each contribution to the self-energy is constructed by solving the corresponding Bethe-Salpeter equations for the electron-hole polarization propagator ($\Sigma^{GW}$), the positron-electron propagator ($\Sigma^{\Gamma}$) and the positron-hole propagator ($\Sigma^{\Lambda}$), see \cite{Hofierka2022} for details. The total self-energy is calculated in this work as the sum of these three contributions: $\Sigma = \Sigma^{GW} + \Sigma^{\Gamma} + \Sigma^{\Lambda}$ (note that in this work $GW$ means $GW$@BSE).

Once the positron wavefunction has been determined, 
the lifetime $\tau$ of the positron-molecule complex against annihilation can be calculated as $\tau = 1 / \pi r_0^2c\delta_{ep}$, where $r_0$ is the classical electron radius and $c$ is the speed of light, so that $\tau$ [ns] = 1 / 50.47$\delta_{ep}$ [a.u.], where the contact density 
\begin{equation}
    \delta_{ep} = \sum_{n=1}^{N_{\rm e}}\gamma_n\int \left| \phi_n(\mathbf{r})\right|^2 \left|\psi_{\varepsilon}(\mathbf{r})\right|^2 d\mathbf{r},
\end{equation}
 in which $\phi_n(\mathbf{r})$ are the $N_{\rm e}$ occupied electronic molecular orbitals,
$\psi_{\varepsilon}(\mathbf{r})$ is the positron wavefunction, and $\gamma_n$ are molecular-orbital-energy-dependent enhancement factors that account for vertex corrections to the electron-positron annihilation vertex that describe short-range electron-positron attractions before annihilation \cite{Green2015}.

\subsection{Implementation via Gaussian bases}
\begin{figure}[h]
    \centering
    \includegraphics[trim={2cm 4cm 0cm 0cm},clip,width=0.8\linewidth]{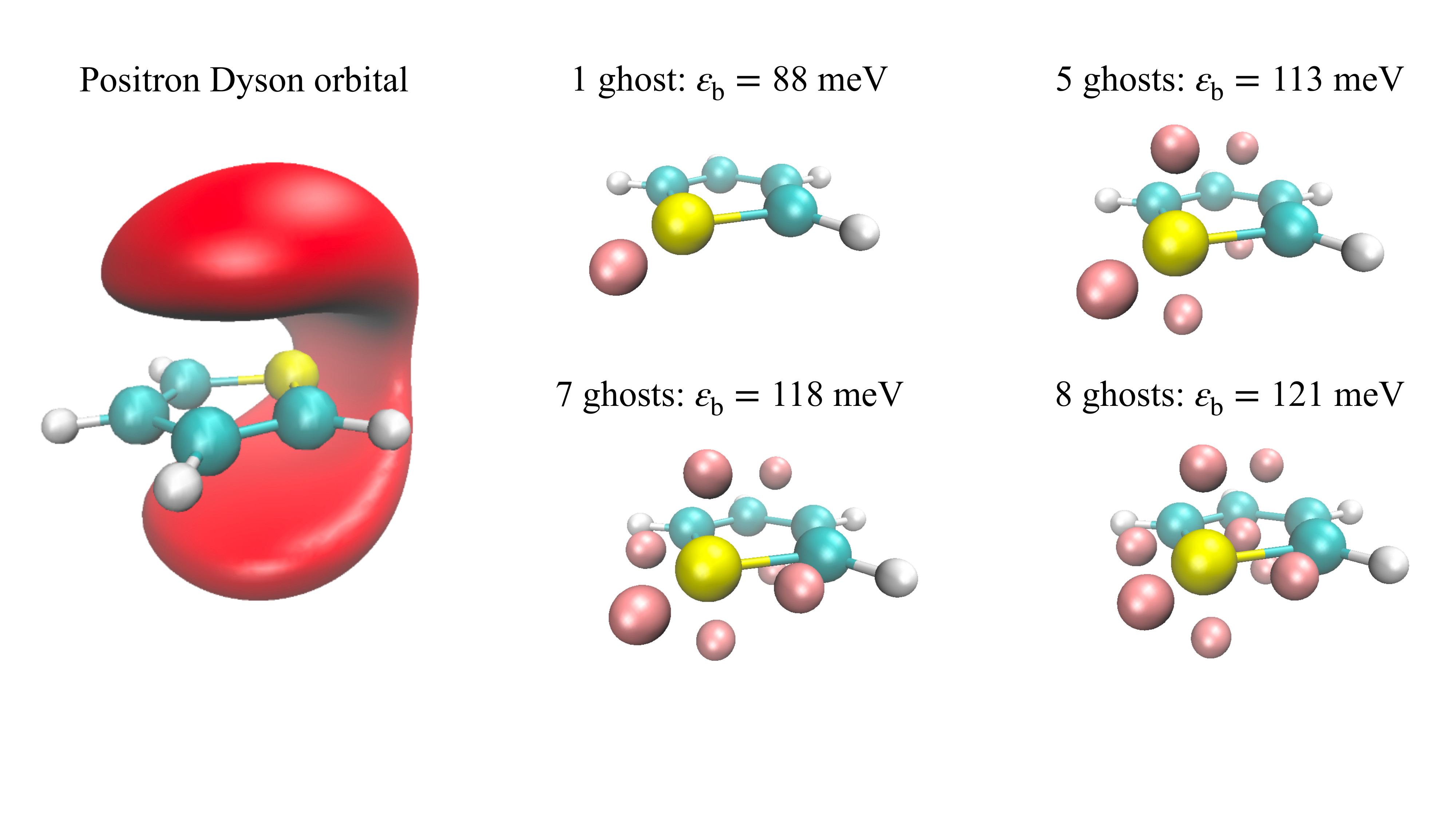}
    \caption{Convergence of the positron binding energy $\varepsilon_{\rm b}$ of thiophene (\ce{C4H4S}) with respect to the number of ghost centers. Diagrams show the thiophene molecule, with green (grey) ((yellow)) spheres for carbon (hydrogen) ((sulfur)) atoms, and selections of ghost centers (pink spheres), labeled with the calculated $\varepsilon_{\rm b}$ at the $GW+\Gamma+\Lambda$ level with bare Coulomb interactions and Hartree-Fock energies in the ladder series. The positron Dyson orbital from the 8-ghost calculation is shown as a red isosurface at 80\% of the maximum value to indicate the region of high positron density. 
    }
    \label{fig:convergence_mo_range}
\end{figure}

We use the fixed-nuclei approximation and optimize molecular geometries at the Hartree-Fock level with aug-cc-pVTZ basis sets \cite{Dunning1992, basis_set_exchange} using the NWChem software \cite{NWChem}. We implement the many-body theory in our open-source {\tt EXCITON+} code, in which the positron and electron wavefunctions are both expanded in (not necessarily equivalent) Gaussian basis sets. 
In the positron-molecule problem one has to account for both long-range interactions (e.g., polarization) and short-range interactions (positron-nuclear repulsion, screening, and virtual-Ps formation). This requires more complicated basis set arrangement than typical all-electron electronic structure calculations. 
First, aug-cc-pV\{T,Q\}Z electron and positron basis sets are placed on all atomic centers; this is required to obtain an accurate representation of the electronic structure and properly describe positron-nuclear repulsion. 
The virtual-Ps formation process takes place ($\sim$1 a.u. or so) away from the molecule. 
To obtain sufficient resolution, additional aug-cc-pV\{T,Q\}Z hydrogen basis sets are then placed on 4--10 chargeless ghost centers placed in region(s) of high positron density (in practice these regions are determined by considering the negative end of the dipole and via preliminary calculations of positron density). Ghost centers have a significant effect on the calculated binding energies. For example, Figure \ref{fig:convergence_mo_range} shows calculated positron binding energies for thiophene (\ce{C4H4S}) with different arrangements of ghost centers. It shows that increasing the number of ghost centers from one to eight increases the positron binding energy by over 30 meV. 
Finally, one must accurately account for the long-range positron-induced polarization of the molecular electron cloud. For this, an additional diffuse even-tempered positron basis is also required on one of the atomic or ghost centers. We use a $10s9p8d7f3g$ even-tempered basis with exponents $\zeta=\zeta_0\beta^{k-1}$, $k=1,2,...,N$, where $N$ is the number of basis functions with a certain angular momentum, $\beta = 2.2$ and $\zeta_0=10^{-3}$ for $s$ and $p$ functions, $\zeta_0=10^{-2}$ for $d$ and $f$ functions, and $\zeta_0=10^{-1}$ for $g$ functions. The positron bound-state wavefunction has the asymptotic form $\psi_{\varepsilon}\sim e^{-\kappa r}/r$ where $\kappa = \sqrt{2\varepsilon_{\rm b}}$ which, to be well-described at distances $r\sim 1/\kappa$, requires a minimum exponent of $\zeta_0 \lesssim \kappa^2 = 2\varepsilon_{\rm b}$ [a.u.]. The broadest Gaussian function we use in this work ($\zeta_0=10^{-3}$) can cover the extent of positron bound states with binding energies above 13.6 meV, which is sufficient for all of the molecules studied here.
Our calculated positron binding energies are converged with respect to the choice of basis such that altering the types of basis or adding more ghosts changes the positron binding energy by no more than 10\%, but in most cases, convergence to within $<$5\% is achieved. 

To summarize, the basis sets used for the present calculations for heterocyclic molecules are as follows: aug-cc-pVTZ positron and electron basis sets on C and H atoms, aug-cc-pVQZ on N, O and S atoms (except for 1,2- and 1,3-oxathiole which have aug-cc-pVTZ on the O atoms). In the region of high positron density, 3--9 ghost centers host aug-cc-pVQZ hydrogen electron and positron bases, 
and at the center of the molecular ring, one ghost center hosts an aug-cc-pVQZ hydrogen electron basis and an even-tempered positron basis as described above. Input files containing geometry and basis information for each molecule are provided in the Supporting Information.

Solving the Dyson equation with the full many-body theory is computationally expensive; in particular, for these heterocycles, obtaining the virtual positronium formation self-energy $\Sigma^{\Gamma}$ required diagonalizing matrices of dimensions between 250,000 $\times$ 250,000 -- 325,000 $\times$ 325,000. The dimension of this matrix is the product of the total number of positron molecular orbitals and excited electron molecular orbitals, so to curb the computational cost, we truncate the number of electron and positron states, typically neglecting positron and electron states with energy above $\sim$150$-$200 eV in the final calculations, and convergence with respect to the number of states included is studied to ensure that the energy denominators in the diagrams are such that saturation occurs. 
Graphs showing the convergence of positron binding energies with respect to the number of states included are available in the Supporting Information. 
In Figure 1, when testing convergence with respect to the number of ghost centers, the calculations were converged to within a few meV with respect to the number of states included, always using at least 70\% of the total positron and electron states, which corresponded to cutoffs of 120$-$160 eV for the electron states and 150$-$175 eV for the positron states.

All of the positron binding calculations were performed using the MPI parallelized {\tt EXCITON+} code running on 10--14 nodes of the UK Tier-2 HPC cluster Kelvin2 \cite{kelvin2} with 640--896 parallel processes in total. The calculations took 12--17 hours to complete, typically requiring 2.0$-$3.5 TB RAM. 
The cost of these calculations motivates using a computationally cheaper model to estimate the virtual positronium formation self-energy $\Sigma^{\Gamma}$. To do this, we scale the bare polarization contribution, $\Sigma^{(2)}$, and construct $\Sigma$ as
\begin{equation}\label{model_self_energy}
    \Sigma \approx g\Sigma^{(2)} + \Sigma^{\Lambda},
\end{equation}
where $g$ is a free parameter. Although this is somewhat unphysical as the asymptotic attraction is overestimated, values of $g=1.4$ and $g=1.5$ often provide good lower and upper bounds of the binding energy \cite{Hofierka2022}.

\section{Predicted positron binding energies of substituted heterocycles}\label{sec:results}

\begin{table*}[ht!]
\centering
\footnotesize
\caption{\label{tab:bind}
Predicted calculated positron binding energies for five-membered heterocycles at the $GW$@BSE and $GW+\Gamma+\Lambda$ levels of many-body theory, and from the model self-energy (Eqn.~ \ref{model_self_energy}). Also shown are the calculated dipole moment $\mu$ (at HF), isotropic polarizability $\alpha$ (at $GW$@BSE level of MBT), ionization energy $I$ (first at HF, second at $GW$@RPA) and contact density (at $GW+\tilde\Gamma+\tilde\Lambda$ with enhancement factors \cite{Green2015}), and the number of $\pi$ bonds. Data for furan and pyrrole are quoted from Ref.~\cite{ArthurBaidoo2024} and are marked with an asterisk (*). Line diagrams of the molecules are given below the table.}

\begin{tabular}{l@{\hskip4pt}l@{\hskip4pt}c@{\hskip4pt}c@{\hskip4pt}c@{\hskip4pt}c@{\hskip4pt}c@{\hskip4pt}c@{\hskip4pt}c@{\hskip4pt}c@{\hskip4pt}}
\\[0.3ex]
\toprule
& & \multicolumn{3}{c}{\textbf{Molecular properties}} & & \multicolumn{3}{c}{\textbf{Positron binding energy (meV)}} & \\ 
\cline{3-5} \cline{7-9}
\\[-1.5ex]
\textbf{Molecule} & \textbf{Formula} & $\mu$ (D) & $\alpha$ (a.u.) & $I$ (eV) & $N_{\pi}$ & $\Sigma^{GW}$ & $\Sigma^{GW+\Gamma+\Lambda}$ $^{[1]}$ & $g\Sigma^{(2)} + \Sigma^{\Lambda}$$^{[2]}$ & $\delta_{ep}$ (a.u.)\\
\hline\\[-1.2ex]
\multicolumn{5}{l}{\textit{Molecules with one heteroatom}} \\[0.5ex]
Furan* & \ce{C4H4O} & 0.7 & 45.4 & 8.85, 9.35 & 2 & $<$0 & 45, 32, \textbf{42} & 20\,--\,49 & 7.32$\times10^{-3}$ \\

Tetrahydrofuran & \ce{C4H8O} & 1.9 & 47.4 & 11.0, 10.1 & 0 & 4.7 & 52, 44, \textbf{53} &   43\,--\,72 & 7.99$\times10^{-3}$ \\

Thiophene & \ce{C4H4S} & 0.8 & 60.6 & 9.08, 9.44 & 2 & $<$0 & 123, 93, \textbf{115} &   \phantom{0}72\,--\,124 & 1.23$\times10^{-2}$ \\

Tetrahydrothiophene & \ce{C4H8S} & 2.3 & 62.5 & 8.95, 8.86 & 0 & 30 & 194, 161, \textbf{189} &   109\,--\,159 & 1.54$\times10^{-2}$ \\

Pyrrole* & \ce{C4H4NH} & 1.9 & 51.1 & 8.22, 8.64 & 2 & 11 & 157, 127, \textbf{148} &  \phantom{0}94\,--\,148 & 1.55$\times10^{-2}$ \\

Pyrrolidine & \ce{C4H8NH} & 1.0 & 51.8 & 9.90, 9.19 & 0 & $<$0 & 46, 35, \textbf{48} &   29\,--\,62 & 9.23$\times10^{-3}$ \\[1.5ex]

\multicolumn{5}{l}{\textit{Molecules with multiple heteroatoms}}\\[0.5ex]
Imidazole & \ce{C3H4N2} & 3.8 & 45.3 & 8.89, 9.34 & 2 & 115 & 289, 263, \textbf{290} &   248\,--\,313 & 1.97$\times10^{-2}$ \\

Pyrazole & \ce{C3H4N2} & 2.3 & 45.3 & 9.53, 9.79 & 2 & 29 & 149, 128, \textbf{147} &   121\,--\,173 & 1.41$\times10^{-2}$ \\

Triazole & \ce{C2H3N3} & 2.9 & 39.6 & 10.5, 10.7 & 2 & 24 & 112, 97, \textbf{113} &   \phantom{0}95\,--\,138 & 1.13$\times10^{-2}$ \\

Oxazole & \ce{C3H3NO} & 1.6 & 40.0 & 9.67, 10.2 & 2 & 0.2 & 65, 53, \textbf{66} &   48\,--\,82 & 8.90$\times10^{-3}$ \\

Isoxazole & \ce{C3H3NO} & 3.2 & 40.2 & 10.3, 10.6 & 2 & 51 & 136, 123, \textbf{136} &   135\,--\,179 & 1.05$\times10^{-2}$ \\

Thiazole & \ce{C3H3NS} & 1.5 & 54.8 & 9.55, 9.94 & 2 & 7.5 & 112, 91, \textbf{109} &   \phantom{0}95\,--\,147 & 1.17$\times10^{-2}$ \\

Isothiazole & \ce{C3H3NS} & 2.7 & 55.0 & 9.69, 9.98 & 2 & 35 & 165, 141, \textbf{162} &   142\,--\,198 & 1.26$\times10^{-2}$ \\

1,3-Oxathiole & \ce{C3H4OS} & 0.6 & 54.0 & 8.14, 8.32 & 1 & $<$0 & 68, 49, \textbf{66} &   29\,--\,62 & 9.84$\times10^{-3}$ \\

1,2-Oxathiole & \ce{C3H4OS} & 2.6 & 54.6 & 8.22, 8.27 & 1 & 22 & 124, 104, \textbf{124} &   \phantom{0}82\,--\,120 & 1.08$\times10^{-2}$ \\[2ex]
\bottomrule
\multicolumn{10}{l}{\includegraphics*[width=0.95\textwidth,trim = {0cm 18cm 2cm 0cm}, clip]{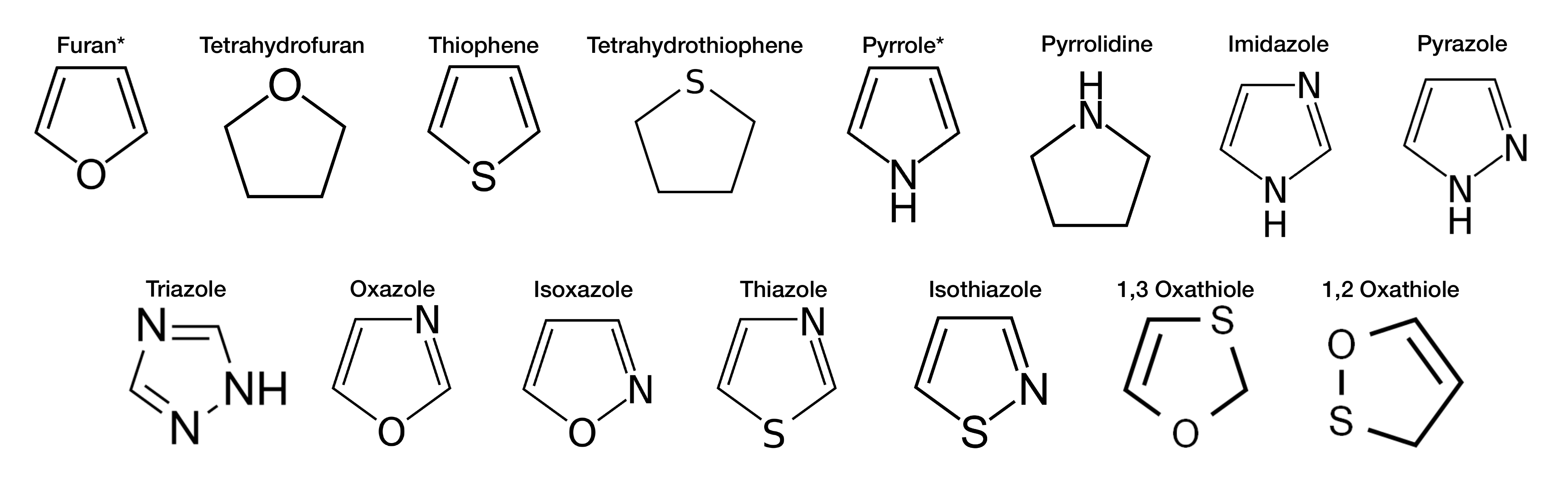}}

\end{tabular}

\begin{flushleft}
[1] Many-body calculations at three levels of $\Sigma^{GW+\Gamma+\Lambda}$: (i) using bare Coulomb interactions within the ladders and HF energies; (ii) using dressed Coulomb interactions within the ladders and HF energies; and (iii) using dressed Coulomb interactions and $GW$@RPA (random phase approximation) energies. The latter (highlighted in bold) is the most sophisticated calculation.

[2] Values of $g=1.4$ and $g=1.5$ are used to give the lower and upper bounds of the range, respectively.

\end{flushleft}
\end{table*}

\begin{figure*}[!ht]
    \centering
    \includegraphics[trim={11cm 1cm 11cm 0cm},clip,width=\textwidth]{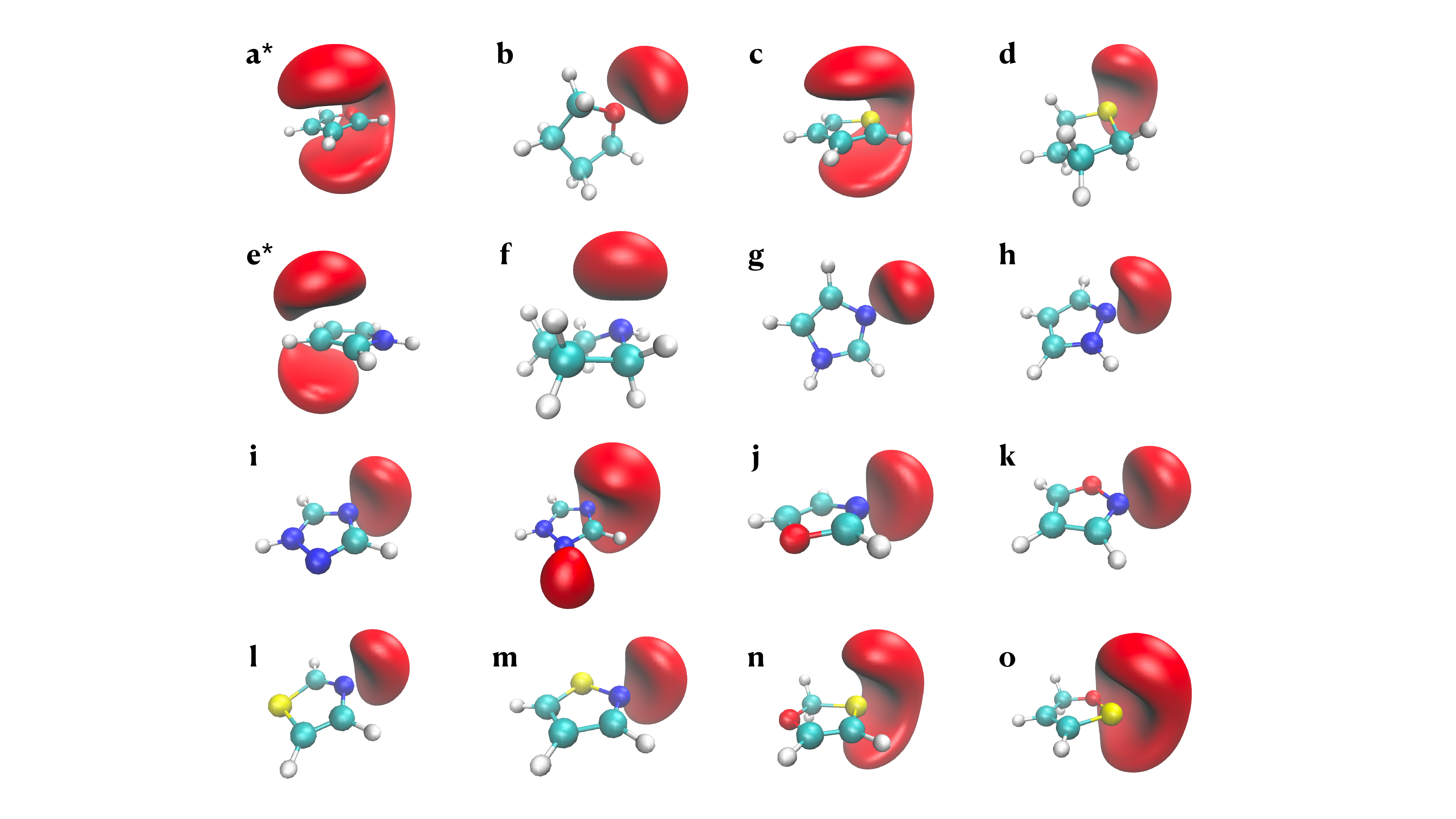}
    \caption{Positron bound-state Dyson orbitals for the five-membered heterocycles in Table \ref{tab:bind}, calculated at the $GW+\tilde\Gamma+\tilde\Lambda$ level of many-body theory.  Dyson orbitals are depicted in red as an isosurface at 80\% of the maximum wavefunction value. Atoms are color-coded as follows: H (grey), C (green), N (blue), O (red), and S (yellow). Molecules shown: (a)* furan, (b) tetrahydrofuran, (c) thiophene, (d) tetrahydrothiophene, (e)* pyrrole, (f) pyrrolidine, (g) imidazole, (h) pyrazole, (i) triazole (80\% isosurface left, 60\% isosurface right), (j) oxazole, (k) isoxazole, (l) thiazole, (m) isothiazole, (n) 1,3-oxathiole, and (o) 1,2-oxathiole. Dyson orbitals for furan and pyrrole are reproduced from \cite{ArthurBaidoo2024} for comparison and marked with an asterisk (*).
    (a)--(f) contain a single heteroatom; the remainder contain two except for (i) that contains three heteroatoms.}
    \label{fig:cyclic_wavefunctions} 
\end{figure*}

Table \ref{tab:bind} presents the predicted positron binding energies for the five-membered heterocycles considered, calculated using three `flavours'
of the full $GW+\Gamma+\Lambda$ many-body theory, which differ by whether bare or screened Coulomb interactions are used in the ladder rungs, and whether HF or $GW$@RPA MO energies are used in diagram construction, as described in the table caption. The final, bold value is from the most sophisticated calculation (which uses screened Coulomb interactions and $GW$@RPA energies) and will hereafter be referred to as the $GW+\tilde\Gamma+\tilde\Lambda$ calculation. To quantify the importance of including the virtual Ps formation process $\Sigma^{\Gamma}$ and positron-hole repulsion $\Sigma^{\Lambda}$ in the self-energy, positron binding energies at the $GW$@BSE level of theory are also given. Results are also given for the much less expensive calculations using the model self-energy of Eqn.~\ref{model_self_energy}.
Also shown in the table are calculated molecular properties including dipole moments $\mu$, polarizabilities $\alpha$ and ionization energies $I$, the number of $\pi$ bonds $N_{\pi}$ in each molecule, and contact densities $\delta_{ep}$. We additionally quote for comparison equivalent calculations by us and colleagues for furan and pyrrole from Ref.~\cite{ArthurBaidoo2024}, which were shown there to be in good agreement with experiment. Figure \ref{fig:cyclic_wavefunctions} contains the corresponding positron bound-state wavefunctions (Dyson orbitals) calculated at the $GW+\tilde\Gamma+\tilde\Lambda$ level, as isosurfaces at 80\% of the maximum value. 
For the most strongly polar molecules the positron bound states are highly localized at the negative end of the molecule. For the weakly polar molecules, the positron bound state is more delocalized and extends across the entire molecule. The effects of heteroatom type and position in the ring are discussed in Section 3.2.1 below.

All the molecules in Table \ref{tab:bind} bind the positron in the full many-body theory, with calculated binding energies in the range 42$-$290 meV in the most sophisticated level of theory. At $GW$@BSE level, only 11 of the 15 molecules bind the positron, with binding energies reaching at most 40\% of the full MBT result. Comparison of the full many-body theory and $GW$@BSE-only results thus highlights the deficiency of the $GW$@BSE diagram alone, and the substantial enhancement that results due to virtual positronium formation (which is slightly mitigated by the positron-hole repulsion \cite{Hofierka2022}).
The model self-energy results [of Eqn.~\ref{model_self_energy}] with scaling factors $g=1.4$ and $1.5$ 
give a (large) range that includes the \emph{ab initio} $GW+\tilde\Gamma+\tilde\Lambda$ results for most of the molecules, but it underestimates binding in three of the sulfur-containing molecules: tetrahydrothiophene (\ce{C4H8S}), 1,3-oxathiole (\ce{C3H4OS}) and 1,2-oxathiole (\ce{C3H4OS}). 

In what follows, we consider correlation between binding energies and molecular properties, and the relationship between electron-positron contact density and binding energy. We then discuss the effect of different types and combinations of heteroatoms and the influence of $\pi$ orbitals and aromaticity on the location and shape of the positron Dyson orbital. Throughout the text, we quote the calculated positron binding energy as the $GW+\tilde\Gamma+\tilde\Lambda$ result from Table \ref{tab:bind} with an associated uncertainty equal to the maximum difference between the three $GW+\Gamma+\Lambda$ results.

\subsection{Polarizability, dipole moment and contact density}

\begin{figure}[h]
    \centering
    \includegraphics[trim={1cm 6cm 0cm 4.5cm},clip,width=\linewidth]{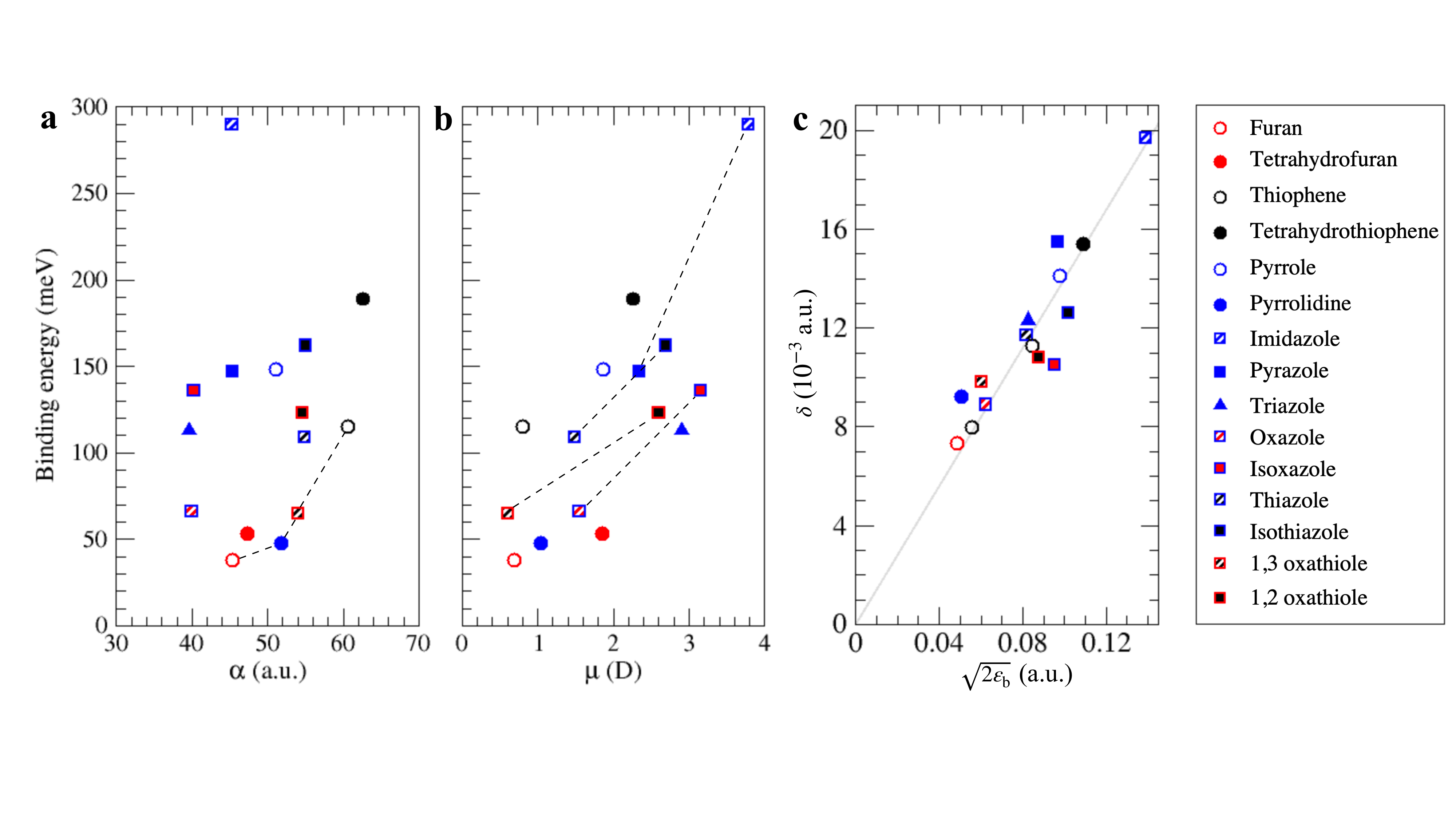}
    \caption{(a) and (b) show positron binding energies (calculated at the $GW+\tilde\Gamma+\tilde\Lambda$ level) vs.~(a) polarizability $\alpha$ and (b) dipole moment $\mu$. Dashed lines in (a) connect the four molecules with the smallest dipole moments of those considered (0.6$-$1~D) to highlight a trend with increasing polarizability and in (b) connect pairs of isomers, illustrating correlation between $\mu$ and $\varepsilon_{\rm b}$ when other properties are almost constant.
    (c) Enhanced positron annihilation contact density 
    plotted against $\sqrt{2\varepsilon_{\rm b}}$ with the line of best fit. }
    \label{fig:molprop}
\end{figure}

Substituted heteroatoms significantly affect molecular properties including the dipole moment and polarizability, which themselves influence positron binding \cite{Danielson22, Danielson09, Danielson25}. In most cases, introducing a heteroatom changes multiple molecular properties simultaneously. For certain subsets, however, significant changes occur in only one property.
Figure \ref{fig:molprop} (a) and (b) show the full many-body theory calculated binding energies vs. isotropic polarizability and dipole moment.
The overall correlation between positron binding energy and polarizability is very weak but some positive correlation is found for smaller sets of molecules with similar dipole moments. For example, there is a positive correlation between the polarizability and binding energy for the weakly polar molecules ($0.6 \leq \mu \leq 1.0$): furan ($\alpha=45.4$ a.u.), pyrrolidine ($\alpha=51.8$ a.u.), 1,3-oxathiole ($\alpha=54.0$ a.u.) and thiophene ($\alpha=60.6$ a.u.). This is highlighted by the dashed line in the figure.
Considering the dipole moment, there is also a large spread in the data. Evidence of correlation is seen more clearly when we consider molecules with similar values of other molecular properties. Within this group of molecules, there are four pairs of isomers (oxazole and isoxazole, thiazole and isothiazole, 1,2- and 1,3-oxathiole, and imidazole and pyrazole) which primarily differ by their dipole moments. Guide lines are included in Figure \ref{fig:molprop}(b) to connect these pairs of isomers. When the effect of changing the dipole moment is isolated in this way, we find in each case that larger dipole moments lead to an increase in the binding energy, as expected. 

In Figure \ref{fig:molprop}(c), we show the contact density vs.~$\sqrt{2\varepsilon_{\rm b}}$. The data align with the known linear relationship $\delta \simeq (F/2\pi) \sqrt{2\varepsilon_{\rm b}}$ \cite{Gribakin2001}, with $F$ = 0.88 here, which 
is similar to but slightly larger than that obtained for atoms ($F$ = 0.66) \cite{Gribakin2001} and halogenated hydrocarbons ($F$ = 0.67) \cite{Cassidy2024}. 

\subsection{Effects of heteroatom locations and substitutions on positron binding} 
The following sections discuss the effect of N, O, S and NH substituents on the positron binding energy and Dyson orbital, considering which heteroatoms are the most attractive to the positron.

\subsubsection{Single-heteroatom systems}
Three molecules in Table \ref{tab:bind} have one distinct heteroatom and no double bonds: tetrahydrofuran (\ce{C4H8O}, $\varepsilon_{\rm b} = 53\pm9$ meV), pyrrolidine (\ce{C4H8NH}, $\varepsilon_{\rm b} = 48\pm13$ meV) and tetrahydrothiophene (\ce{C4H8S}, $\varepsilon_{\rm b} = 189\pm33$ meV). Although these differ only by the type of heteroatom, their positron binding energies vary by more than a factor of three, with the sulfur heteroatom leading to the largest binding energy.
Figure \ref{fig:cyclic_wavefunctions} (b), (d) and (f) show that for each of these molecules, the positron wavefunction is localized to the region next to the heteroatom which corresponds to the negative end of the molecular dipole. 

Furan (\ce{C4H4O}, $\varepsilon_{\rm b} = 42\pm10$ meV), pyrrole (\ce{C4H4NH}, $\varepsilon_{\rm b} = 148\pm30$ meV) and thiophene (\ce{C4H4S}, $\varepsilon_{\rm b} = 115\pm30$ meV) also have one distinct heteroatom, but differ from the first set of molecules as they are planar, giving them higher symmetry, and contain double bonds in the ring. Again, the positron binding energies differ by more than a factor of three for these molecules, and the sulfur heteroatom leads to the largest binding energy. 
Substituting O with NH has a much greater effect on the binding energy here than in the first case. In tetrahydrofuran and pyrrolidine, the positron bound states are qualitatively similar: the negative end of the molecular dipole moment is at the heteroatom, and the positron binds near this region in both systems. However from furan to pyrrole, the character of the positron bound state changes more significantly: on swapping O to NH, the molecular dipole becomes much stronger and reverses direction and consequently, the positron bound state is altered. In furan, the positron is delocalized across the molecular plane with some concentration at the heteroatom (which corresponds to the negative end of the molecular dipole), whereas in pyrrole it is instead concentrated in two smaller regions above and below the molecular plane, at the opposite side to the heteroatom. This change leads to stronger positron binding in pyrrole than furan.

Double ($\pi$) bonds give rise to $\pi$-type molecular orbitals whose electron density characteristically has a node in the plane of the molecule. Much of their electron density is in the regions above and below the molecular plane, farther from the atomic centers than in other molecular orbitals, and is easier for the positron to probe. In this way, $\pi$ molecular orbitals are expected to draw the positron wavefunction into the regions above and below the molecular plane. This influence is evident in the positron wavefunctions in Figure (a)* furan, (c) thiophene and (e)* pyrrole: the Dyson orbitals are delocalized with regions above and below the plane of the molecule. 
In these wavefunctions, we see evidence of competition between the negative end of each molecule and the $\pi$ bonds, both of which attract the positron.
 
The presence of $\pi$ bonds in these single-heteroatom molecules does not have a uniform effect on the positron binding energy: when the $\pi$ bonds in furan and thiophene are removed to form tetrahydrofuran and tetrahydrothiophene, the positron binding energy increases, but removing the $\pi$ bonds in pyrrole to form pyrrolidine decreases the binding energy. 
In furan and thiophene, removing the $\pi$ bonds reinforces the dipole moment, providing a stronger negatively charged region for the positron to attach to, while in pyrrole, the dipole moment reverses direction and becomes weaker when the $\pi$ bonds are removed, resulting in less favorable conditions for positron binding.

\subsubsection{Multiple-heteroatom systems}
Figure \ref{fig:cyclic_wavefunctions} (g--o) contains positron wavefunctions for molecules with multiple heteroatoms. 
All these molecules are aromatic except for 1,2- and 1,3-oxathiole [Fig.~2 (n) and (o)] which have a fully saturated carbon atom disrupting the conjugation of the ring.
We assess how N, O, S and NH substituents in the ring affect the positron orbital location and ascertain which substituents the positron is most and least attracted to. 

\begin{figure}[h]
    \centering
    \includegraphics[trim={0cm 27cm 0cm 1cm}, clip,width=\linewidth]{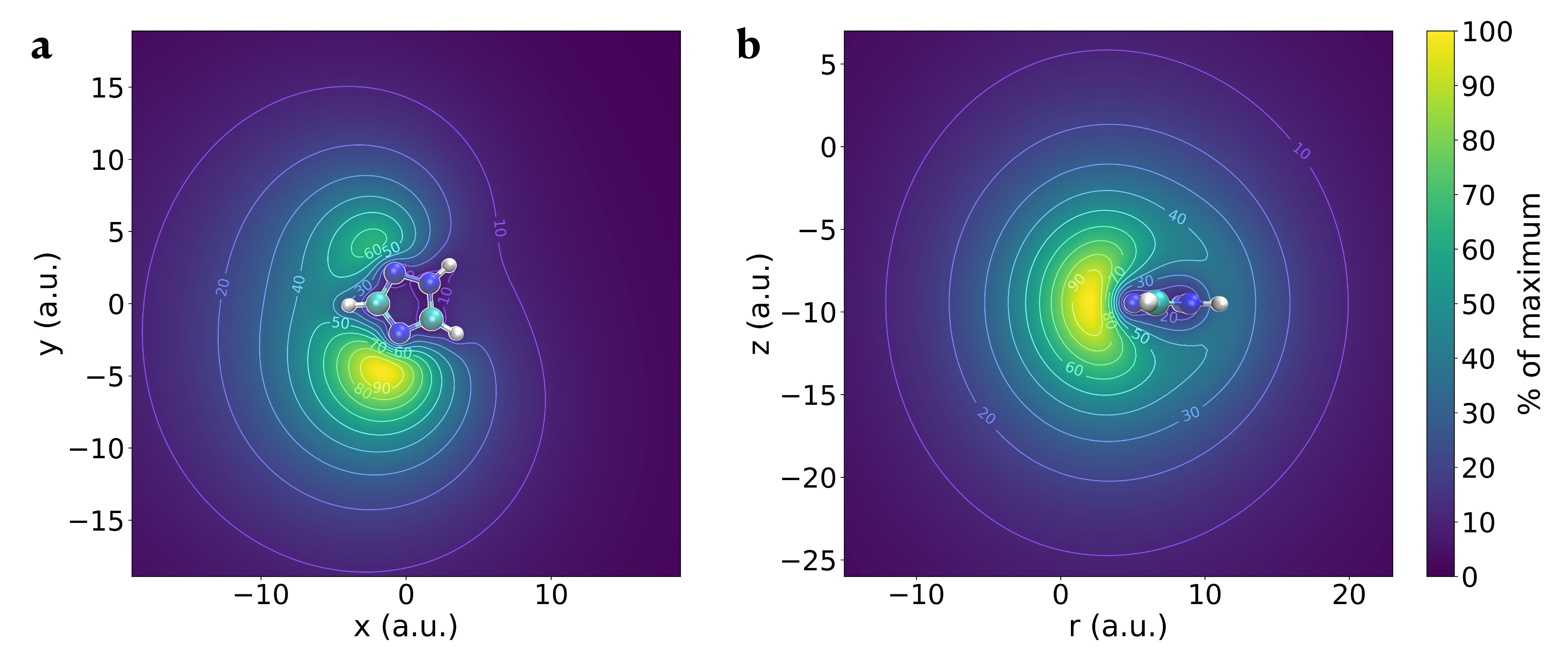}
    \caption{The positron bound-state wavefunction for triazole ($\ce{C2H3N3}$) (a) in the molecular plane and (b) in a plane perpendicular to the molecule through the line from the N atom with the highest positron density to the midpoint of the other two N atoms. Contours are shown from 10\% to 90\% of the maximum value in 10\% increments. 
    The ball-and-stick image shows carbon as green, nitrogen as blue and hydrogen as white.} 
    \label{fig:triazole2D}
\end{figure}

Figure \ref{fig:cyclic_wavefunctions} (g--i) shows that the positron is more attracted to the N atoms over NH groups in imidazole, pyrazole and triazole. This can be explained by noting that the additional H atom in the NH group makes the electron density of the N atom difficult to probe as it is involved in binding to the H atom, in contrast to the N substituent which is instead involved in a double bond with the neighboring C atom, thus having some of its electrons in $\pi$ molecular orbitals. Triazole, the only molecule with three heteroatoms, is also the only one with a distinct secondary lobe of high positron density. The secondary lobe is not visible in the 80\%-of-maximum isosurface, so we also show a second image with an isosurface at 60\% of the maximum in Fig.~\ref{fig:cyclic_wavefunctions} (i, right). Moreover, to examine the positron density in more detail, Fig.~\ref{fig:triazole2D} shows 2D contour plots of the positron Dyson orbital in the molecular plane (a) and in a plane perpendicular to the molecule (b). 
In the molecular plane, there are two distinct lobes of high positron density; the primary lobe is localized next to the N heteroatom that is bonded to both carbons and aligned with the negative end of the molecular dipole moment, while the secondary lobe is at the opposite side of the ring, next to another N heteroatom, where significant electron density must exist. 
In the plane perpendicular to the molecule, the positron density is drawn into the regions above and below the ring, particularly visible in the 40\% contour.  This indicates some influence of electrons in $\pi$ orbitals on the bound-state wavefunction, despite the large dipole moment ($\mu = 2.9$ D). 

Figure \ref{fig:cyclic_wavefunctions} (j--m) show that the positron is more attracted to the N atoms than the O atoms in oxazole and isoxazole, and the S atoms in thiazole and isothiazole. We can therefore conclude that of the four types of substituent, N is the most attractive to the positron. Now, from the positron wavefunctions of 1,2- and 1,3-oxathiole in Figure \ref{fig:cyclic_wavefunctions} (n--o), we see a slight preference of the positron to attach to the S atom over the O atom. In all cases, substituting an O atom for an S atom increases the positron binding energy (comparing e.g., isoxazole to isothiazole one sees that the latter has a larger polarizability and smaller ionization energy). 
N, O and S substituents are all more attractive to the positron than the C atoms remaining in the ring, but the NH group is the least attractive. The wavefunction for the singly-substituted molecule pyrrole in Figure \ref{fig:cyclic_wavefunctions}(e)* shows this most clearly with the positron wavefunction density being mostly at the opposite side of the molecule to the NH group, attracted to the double bonds between the carbon atoms. However when the molecule is fully saturated, like pyrrolidine in Figure \ref{fig:cyclic_wavefunctions}(f), the positron prefers to bind next to the NH group rather than the single-bonded carbon atoms.
We can therefore conclude that the positron is most attracted to N substituents, followed by S, O, then NH. 

The relative locations of the heteroatoms in the ring are also important for positron binding: when attractive heteroatoms are next to each other in the ring, this typically leads to a larger dipole moment and stronger binding overall, e.g., there is stronger binding in isothiazole (\ce{C3H3NS}, $\varepsilon_{\rm b} = 162\pm24$ meV) with the N and S atoms next to each other in the ring than in thiazole (\ce{C3H3NS}, $\varepsilon_{\rm b} = 109\pm21$ meV), where they are separated by a C atom. 
The same trend is seen for two other pairs of molecules: isoxazole (\ce{C3H3NO}, $\varepsilon_{\rm b} = 136\pm13$ meV) and oxazole (\ce{C3H3NO}, $\varepsilon_{\rm b} = 66\pm13$ meV), and 1,2-oxathiole (\ce{C3H4OS}, $\varepsilon_{\rm b} = 124\pm20$ meV) and 1,3-oxathiole (\ce{C3H4OS}, $\varepsilon_{\rm b} = 66\pm19$ meV). 
However, the opposite is true when one of the substituents is the NH group; in this case, the dipole moment and positron binding energy both decrease when the heteroatoms are placed closer together, as seen in pyrazole (\ce{C3H4N2}, $\varepsilon_{\rm b} = 147\pm21$ meV) and imidazole (\ce{C3H4N2}, $\varepsilon_{\rm b} = 290\pm27$ meV).

\subsection{Aromaticity and \texorpdfstring{$\pi$}{pi} orbitals}
All but three of the molecules in this work have $\pi$ electrons, characterized by molecular orbitals with a nodal plane through the nuclei and typically arising from double bonds or lone pairs on heteroatoms.
The role of $\pi$ bonds in enhancing positron binding, owing to the ability of the positron to access electron density in regions away from repulsive nuclei, was first suggested in analysis of measurements \cite{Danielson09}, subsequent model potential calculations \cite{Tachikawa:MCP_a,Ashiba2025}, and confirmed and quantified via \emph{ab initio} many-body theory calculations for a small selection of molecules in \cite{Hofierka2022,ArthurBaidoo2024}. 
In this section, we consider further the role of aromaticity and $\pi$ electrons on positron binding in the heterocycles. 

\subsubsection{Aromaticity}
Ten of the molecules in Table \ref{tab:bind} are aromatic, having six $\pi$ electrons which form a fully conjugated ring, namely furan, thiophene, pyrrole, imidazole, pyrazole, triazole, oxazole, isoxazole, thiazole and isothiazole. Each of these has two double bonds providing four $\pi$ electrons, and a third pair of $\pi$ electrons from a lone pair on one of the heteroatoms that is donated to the conjugated system. 

\begin{figure}
    \centering
    \includegraphics[width=0.9\linewidth]{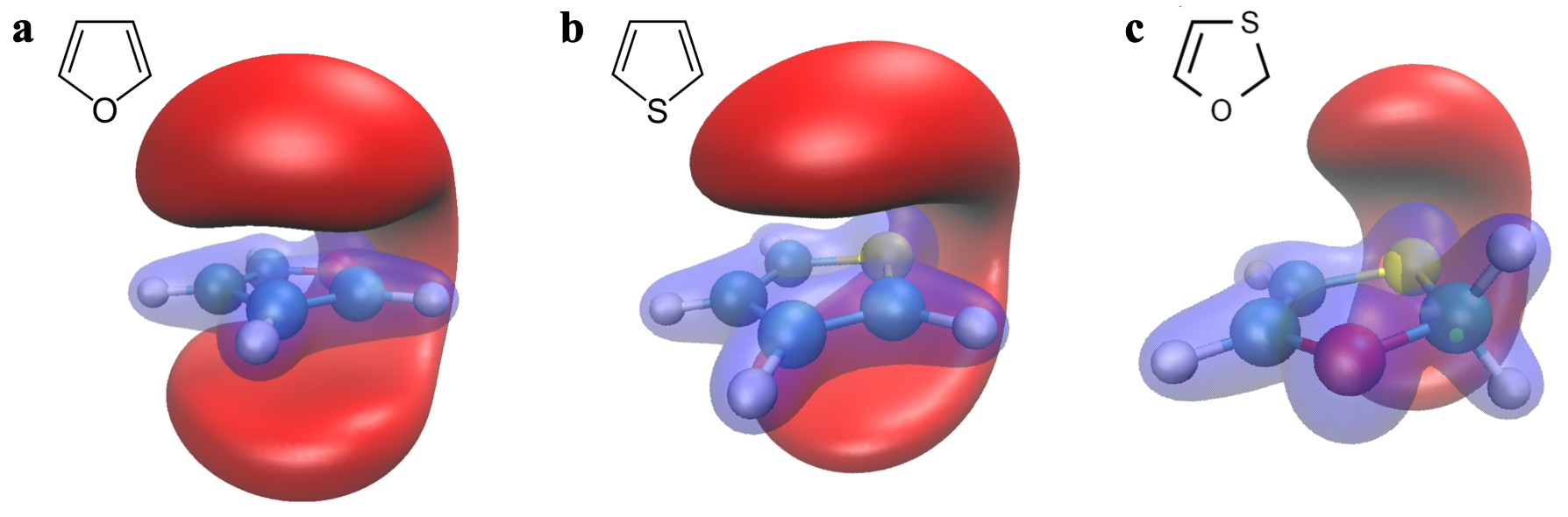}
    \caption{Total electron density from the 10 HOMOs as a transparent blue isosurface at a value of 0.04 a.u., and the positron Dyson orbital as a solid red isosurface at 80\% of the maximum value for (a) furan (\ce{C4H4O}), (b) thiophene (\ce{C4H4S}) and (c) 1,3-oxathiole (\ce{C3H4OS}).}
    \label{fig:aromaticity}
\end{figure}

Figure \ref{fig:aromaticity} shows the total electron density from the ten HOMOs of furan, thiophene and 1,3-oxathiole (which include all the $\pi$ orbitals) as a blue isosurface at a value of 0.04 a.u.. The aromatic furan and thiophene rings exhibit a more even distribution of electron density than non-aromatic 1,3-oxathiole, which has larger lobes of electron density localized at the O and S heteroatoms than the C atoms. This difference is reflected in the positron wavefunctions: for both furan and thiophene, much of the positron density is drawn into the regions above and below the ring but for 1,3-oxathiole the positron density is more localized toward the negative end of the molecular dipole, with evidence of some attraction to the $\pi$ orbitals at the negative side of the molecule but not to the full region above and below the ring.

For aromatic molecules containing multiple heteroatoms, their relative positions also affect aromaticity: placing two heteroatoms adjacent in the ring reduces aromaticity, decreasing $\pi$-electron delocalization above and below the ring. As a result of this reduced aromaticity (and often, also an increased dipole moment), the positron Dyson orbitals are more strongly localized to attractive heteroatoms when they are adjacent in the ring (see, for example, Figure \ref{fig:cyclic_wavefunctions} (j) oxazole and (k) isoxazole, where the latter has a more strongly localized positron orbital).

\subsubsection{Molecular orbital contributions to the self-energy}
Our previous work \cite{Hofierka2022,Cassidy2024,ArthurBaidoo2024}  and Fig.~3 above have revealed that the global molecular properties of dipole moment, isotropic polarizability and ionization energy are insufficient to explain trends in positron binding energies. A strength of our many-body theory approach is the capability to quantify contributions to the positron-molecule correlation potential from individual electron molecular orbitals (MOs).
The total strength of the correlation potential can be quantified using strength parameters $S$ \cite{Dzuba1994}, calculated from diagonal matrix elements of the self energy as
\begin{equation}
    S = -\sum_{\nu} \frac{\langle \nu | \hat{\Sigma}_{\varepsilon} |\nu \rangle}{\varepsilon_{\nu}},
\end{equation}
where $\nu$ is an excited positron Hartree-Fock orbital with energy $\varepsilon_{\nu}$. In this work we take $\Sigma = \Sigma^{GW}+\Sigma^{\Gamma}+\Sigma^{\Lambda}$ and we can consider the strength of individual contributions e.g., the partial strength from the virtual-Ps contribution $S^{(\Gamma)}$. 
Moreover, for the second-order polarization  $\Sigma^{(2)}$ and the virtual-Ps diagrams $\Sigma^{(\Gamma)}$, which both involve a single hole line and include sums over the holes (occupied electrons), one can consider the partial contribution $S_n$ from individual molecular orbitals $n$, e.g., $S^{(\Gamma)}=\sum_n S^{(\Gamma)}_n$. Doing so can provide crucial insight. 
The relative contributions to the self-energy from bare polarization and virtual positronium formation are obtained as the ratio $S^{(2+\Gamma)} / S^{(2)}$ (where $S^{(2)} = S^{(2+\Gamma)} - S^{(\Gamma)}$), with typical values ranging from $\sim$1--1.5. 

Strength parameters typically decrease with the ionization energy of the MO: more tightly-bound MOs are less easily perturbed by the positron, and are also less effectively able to tunnel to the positron to form virtual-Ps; core molecular orbitals make a near-zero contribution. The density of MOs in a given energy range is thus a key quantity beyond simply the ionization energy. However, MO contributions to the correlation potential are not strictly ordered by their energy; the shape and location of the MO relative to the high positron density also play an important role \cite{Hofierka2022}. 

\begin{figure}[t!]
    \centering
    \includegraphics[trim={0 1cm 14cm 0},clip,width=\linewidth]{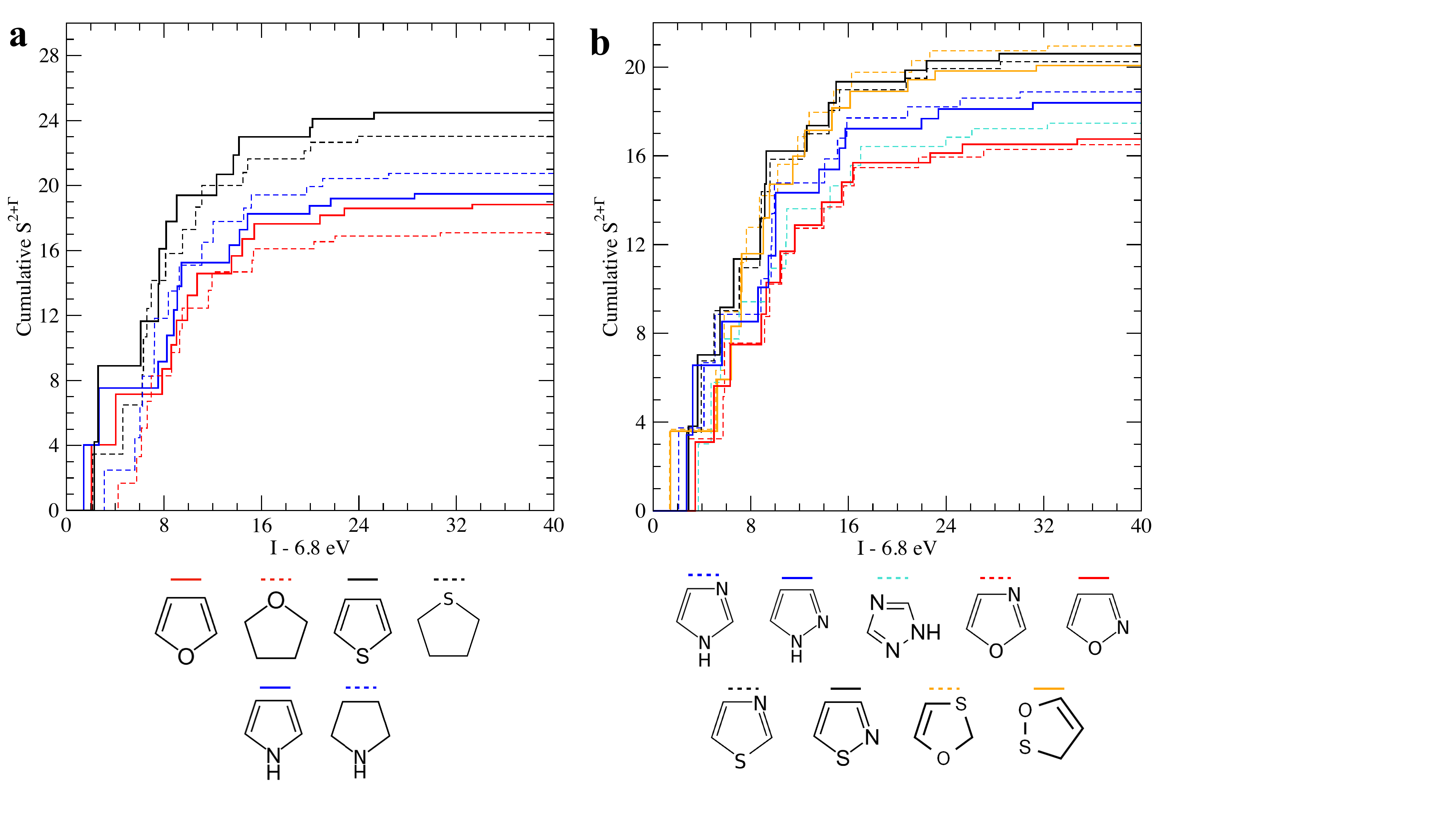}
    \caption{Cumulative contribution from individual MOs to the total positron-molecule correlation potential strength parameter $S^{(2+\Gamma)}$ calculated at the $GW+\tilde\Gamma+\tilde\Lambda$ level of theory for (a) single-heteroatom systems and (b) multi-heteroatom systems.}
    \label{fig:str_param_all}
\end{figure}

Figure \ref{fig:str_param_all} contains cumulative strength parameters $S^{(2+\Gamma)}$ calculated at $GW+\tilde\Gamma+\tilde\Lambda$ level providing an overview of the strength of the positron-molecule correlation potentials for the molecules in Table \ref{tab:bind}. (When interpreting total binding energies, keep in mind that this quantity does not capture the static interaction from the dipole).   From (a), we find that for the single-heteroatom systems, the largest cumulative $S^{(2+\Gamma)}$ (and thus, the most attractive positron-molecule correlation potential) is obtained with sulfur substitutents, in thiophene (black solid line) and tetrahydrothiophene (black dashed line), and the least attractive potentials are obtained with oxygen substitutents, in furan (red solid line) and tetrahydrofuran (red dashed line). This helps to explain, for example, the fact that the calculated binding energy of furan ($\varepsilon_{\rm b} = 42\pm13$ meV) is almost three times smaller than that of thiophene ($\varepsilon_{\rm b} = 115\pm30$ meV), despite their similar structures and dipole moments. 
The strength parameters for multi-heteroatom systems in (b) similarly aid interpretation of the binding energies. For example, isoxazole (red solid line, $\varepsilon_{\rm b} = 136\pm13$ meV) has a smaller binding energy than isothiazole (black solid line, $\varepsilon_{\rm b} = 162\pm24$ meV), although it has a notably larger dipole moment. Strength parameters indicate that isothiazole has a considerably more attractive correlation potential than isoxazole, and thus, explain this difference.
Another example is found in oxazole (red dashed line, $\varepsilon_{\rm b} = 66\pm13$ meV) and 1,3-oxathiole (orange dashed line, $\varepsilon_{\rm b} = 66\pm19$ meV). These have the same calculated positron binding energy despite oxazole having a larger dipole moment by 1 D, but the difference in the attraction to the dipoles is compensated by 1,3-oxathiole having a much more attractive correlation potential.

\begin{figure*}[t!]
    \centering
    \includegraphics[trim={12cm 1cm 12cm 1cm},clip,width=\textwidth]{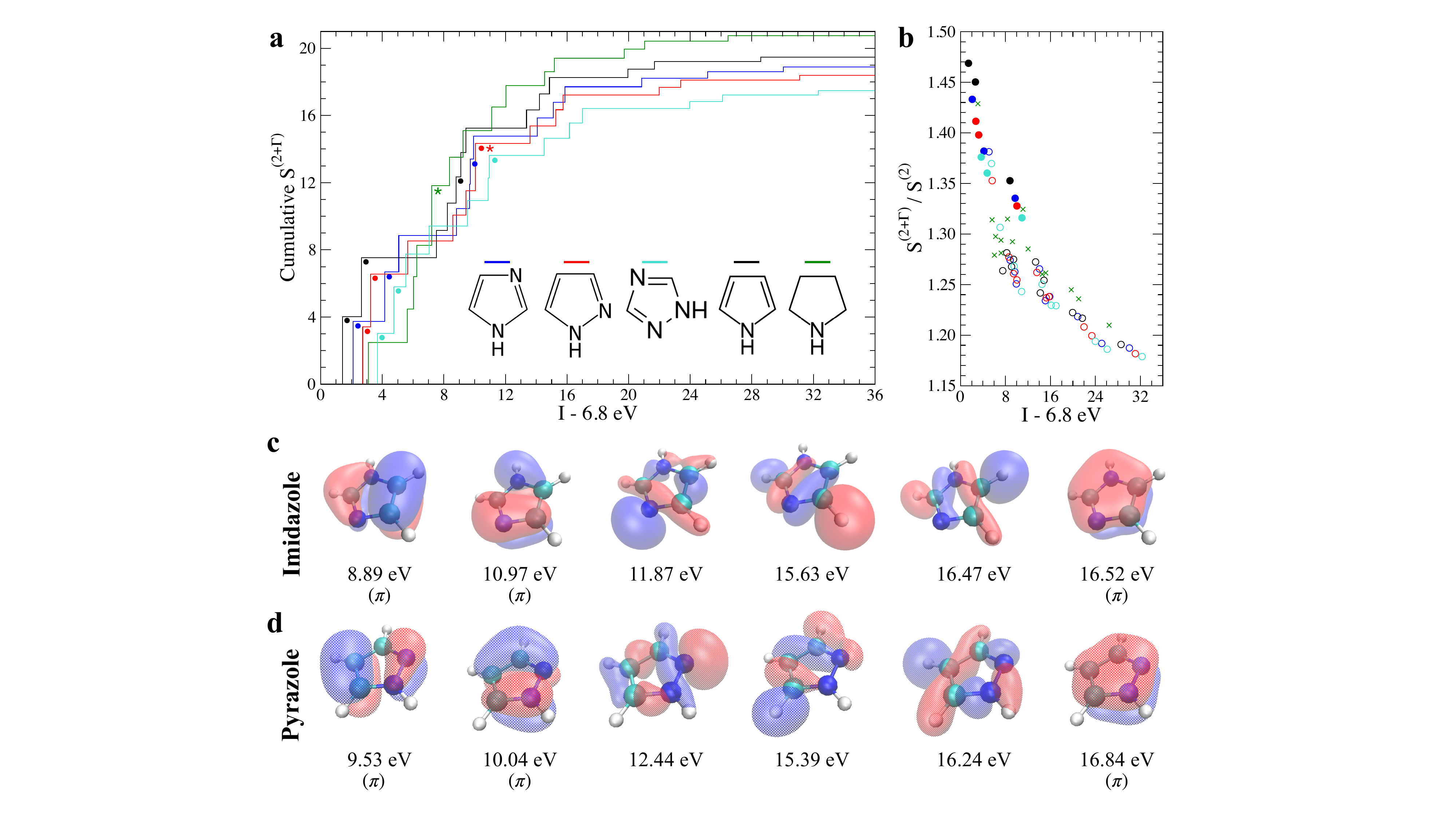}
    \caption{(a) Cumulative contribution from individual MOs to the total positron-molecule correlation potential strength parameter $S^{(2+\Gamma)}$, and (b) values of $S^{(2+\Gamma)} / S^{(2)}$ for imidazole (blue), pyrazole (red), triazole (turquoise), pyrrole (black) and pyrrolidine (green), plotted against MO energy $I-$6.8 eV, where 6.8 eV is the Ps ground state energy. Solid circles denote $\pi$-type orbitals in both graphs, and stars in (a) mark near-degenerate orbitals that are indistinguishable on the graph. The six HOMOs of (c) imidazole and (d) pyrazole are shown below the graphs to illustrate their shapes, labeled with their ionization energies and plotted as red and blue transparent isosurfaces at values of $\pm$0.04 a.u.}
    \label{fig:str_param}
\end{figure*} 

In what follows, we focus on the strength parameters of $\pi$ molecular orbitals: due to their shape, with a node in the plane of the molecule, these molecular orbitals have a more positron-accessible electron density than others. Thus, they are expected to make an enhanced contribution to the self-energy (and thus, to have larger strength parameters) compared with other MOs of similar ionization energy. Such orbitals are found in all but three of the molecules in this work: each of the aromatic molecules has three $\pi$ molecular orbitals, and the oxathioles have four. In the aromatic molecules, the HOMO and HOMO--1 are $\pi$ orbitals, and the third $\pi$ orbital is more deeply bound, typically lying between the HOMO--3 and HOMO--6, depending on the molecule. 
The oxathioles both have $\pi$-type HOMO, HOMO--1, HOMO--4 and HOMO--7, but are not aromatic so the $\pi$ orbitals are each more localized to a particular region in the molecule. Figure \ref{fig:str_param} shows (a) cumulative strength parameters $S^{(2+\Gamma)}$ and (b) values of the ratio $S^{(2+\Gamma)} / S^{(2)}$ for the MOs of molecules in Table \ref{tab:bind} with combinations of N and NH substituents, and images of the six HOMOs of (c) imidazole and (d) pyrazole with the $\pi$ orbitals labeled. The features of these strength parameter plots discussed below also appear in equivalent plots for the other molecules, but the data are omitted from the graphs for clarity. Of the molecules in Figure \ref{fig:str_param}, imidazole, pyrazole, triazole and pyrrole are aromatic, and pyrrolidine is non-aromatic, having no $\pi$ orbitals.

Strength parameters are typically largest for the HOMOs due to their low ionization energy, but when the HOMOs are of $\pi$ character, the difference between their strength parameters and those of the next-most-tightly-bound orbitals is more pronounced. This is seen in Figure \ref{fig:str_param}(a), where the cumulative contribution of individual MOs to the total strength parameter rises rapidly for the $\pi$-type HOMO and HOMO$-$1 of imidazole, pyrazole, triazole and pyrrole, then slows down for the remaining orbitals. However for pyrrolidine, with no $\pi$ orbitals, the cumulative strength parameter rises at a more steady pace, indicating more uniform contributions to the self-energy from the MOs, and ultimately resulting in the strongest correlation potential. 
Overall this highlights how the coarse molecular properties such as ionization energy are alone insufficient to predict the binding energies, and thus the importance of \emph{ab initio} calculations that include a substantial number of MOs and account for their symmetry (in the static interaction and dynamical correlations).

The most deeply-bound $\pi$ orbital in these aromatic molecules sometimes exhibits a larger strength parameter than MOs with comparable ionization energies. However, this orbital is more clearly identified by its high relative contribution to the virtual positronium formation self-energy $\Sigma^{\Gamma}$ compared to the bare polarization term $\Sigma^{(2)}$. This leads to the ratio of strength parameters $S^{(2+\Gamma)} / S^{(2)}$ being higher for this molecular orbital than neighboring molecular orbitals with similar ionization energies. Figure \ref{fig:str_param}(b) demonstrates this clearly: for each of the four aromatic molecules, there is a cluster of points just before and after the $\pi$ orbital with smaller values of $S^{(2+\Gamma)} / S^{(2)}$, and a sudden jump to a larger value for the $\pi$ orbital.  In contrast, for pyrrolidine without the $\pi$ orbital, the values of $S^{(2+\Gamma)} / S^{(2)}$ form a relatively uniform cluster, without evidence of any such structure.
For all of the aromatic molecules studied, the ratio $S^{(2+\Gamma)} / S^{(2)}$ is consistently larger for the most deeply-bound $\pi$ orbital than for neighboring molecular orbitals.

\section{Summary}\label{sec:conclusion}
We have presented the first predictions of positron binding energies and bound-state wavefunctions for a chemically diverse set of five-membered heterocycles containing C, H, N, O, and S. Notably, these include the first many-body theory calculations of positron binding for sulfur-containing molecules.  The calculations were performed using \textit{ab initio} many-body theory where the positron bound state energies and wavefunctions are obtained by solving the Dyson equation, accounting for the dominant electron-positron correlations including positron-induced polarization and screening treated at the GW@BSE level, virtual positronium formation through the infinite electron-positron ladder series, and the corresponding positron-hole repulsion.
At the most sophisticated level of theory, all 13 molecules considered are predicted to bind the positron, with binding energies in the range 42--290 meV. 

Beyond establishing binding energies, the calculations reveal relationships between the structure and properties of the substituted heterocycles and positron binding. Analysis of the positron Dyson orbitals shows a consistent hierarchy of heteroatom attractiveness, N > S > O > NH, while $\pi$ bonds and aromaticity promote delocalization of the positron density above and below the molecular plane. Plots of the Dyson orbitals illustrate competition between these effects for the localization of the wavefunction, in agreement with our previous work \cite{ArthurBaidoo2024}. The interplay of these effects depends sensitively on heteroatom species and relative placement within the ring, with adjacent heteroatoms mainly (but not always) leading to stronger binding through enhanced polarity and electron density accessible to the positron.

Analysis of molecular orbital contributions to the positron-molecule correlation potential highlights that aromatic $\pi$ orbitals make distinct contributions to the correlation potential: $\pi$-type HOMOs make above-average contributions to the correlation potential, while the most deeply bound $\pi$ molecular orbital in an aromatic ring exhibits a characteristically large relative contribution to the virtual positronium formation part of the positron-molecule self-energy than orbitals of similar ionization energy. These trends provide a useful framework for interpreting positron-molecule interactions in aromatic molecules, and underscore that positron binding energies cannot be reliably inferred from global molecular descriptors alone and instead depend critically on orbital symmetry, spatial distribution, and energetic alignment. 

From a theoretical perspective, this work significantly extends the range of molecular systems to which the recently developed \emph{ab initio} many-body theory of positron binding has been successfully applied, demonstrating its predictive capability for chemically nontrivial heterocyclic molecules. The work has laid a foundation for the use of these many-body theory methods to study larger and more complex molecules, particularly those of biological relevance such as amino acids, radiotracers for PET and DNA nucleobases, of which some of the heterocycles studied here are components.  
The results reported here provide benchmark data for alternative approaches to the positron-molecule problem, which is a challenging testbed for theoretical and computational approaches to the many-fermion problem, and for emerging data-driven approaches trained on many-body-theory reference results (see, e.g., \cite{Cassella2024}). 
More broadly, a quantitative understanding of positron binding energies represents a necessary first step toward a complete description of vibrational Feshbach resonances and underpins the developing use of low-energy positrons as a novel and chemically sensitive form of molecular spectroscopy.

\section*{Acknowledgements}
We thank Jaroslav Hofierka, James Danielson, Eugene Arthur Baidoo, Gleb Gribakin, and the late Cliff Surko for useful discussions, and Brian Cunningham, Steven Cousens and Northern Ireland HPC for high-performance computing support.\\ 

\noindent We acknowledge support from the European Research Council, grant agreement No. 101170577 (D.G.G).\\

\noindent Images containing ball-and-stick representations of molecules and wavefunction isosurfaces were generated using VMD \cite{VMD1996}.

\section*{Supporting information}
 
 
\begin{itemize}
\item Convergence and Basis Sets: Graphs showing convergence of positron binding energies with respect to the number of excited electron and positron molecular orbitals included in the calculation of the self-energy diagrams, and a list of basis sets used in the calculations in Table \ref{tab:bind} (PDF).
\item INPUTs: Geometry and basis input files for the {\tt EXCITON+} code used to obtain the positron bound states in Table \ref{tab:bind} (ZIP).
\end{itemize}

\includepdf[pages=-]{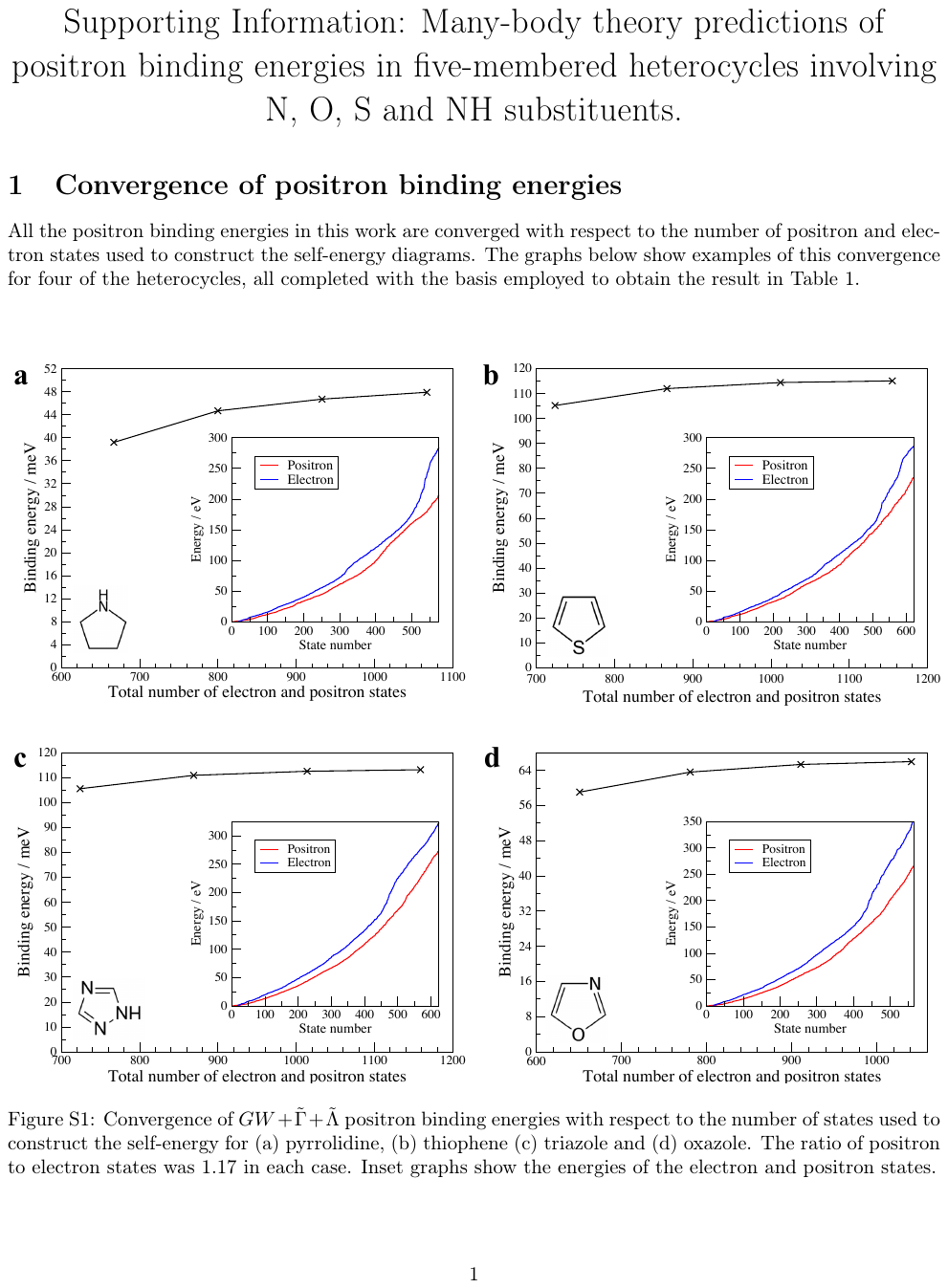}

\printbibliography

@article{Kelly1963,
  title = {Correlation Effects in Atoms},
  author = {Kelly, Hugh P.},
  journal = {Phys. Rev.},
  volume = {131},
  issue = {2},
  pages = {684},
  numpages = {0},
  year = {1963},
  month = {7},
  publisher = {American Physical Society},
  doi = {10.1103/PhysRev.131.684},
  url = {https://link.aps.org/doi/10.1103/PhysRev.131.684}
}

@article{Kelly1964,
  title = {Many-Body Perturbation Theory Applied to Atoms},
  author = {Kelly, Hugh P.},
  journal = {Phys. Rev.},
  volume = {136},
  issue = {3B},
  pages = {B896},
  numpages = {0},
  year = {1964},
  month = {11},
  publisher = {American Physical Society},
  doi = {10.1103/PhysRev.136.B896},
  url = {https://link.aps.org/doi/10.1103/PhysRev.136.B896}
}

@article{Kelly1968,
  title = {Many-Body Perturbation Theory Applied to Open-Shell Atoms},
  author = {Kelly, Hugh P.},
  journal = {Phys. Rev.},
  volume = {144},
  issue = {1},
  pages = {39},
  numpages = {0},
  year = {1966},
  month = {4},
  publisher = {American Physical Society},
  doi = {10.1103/PhysRev.144.39},
  url = {https://link.aps.org/doi/10.1103/PhysRev.144.39}
}

@article{Crawford1967,
doi = {10.1088/0370-1328/91/2/303},
url = {https://doi.org/10.1088/0370-1328/91/2/303},
year = {1967},
month = {6},
publisher = {},
volume = {91},
number = {2},
pages = {279},
author = {O H Crawford},
title = {Bound states of a charged particle in a dipole field},
journal = {Proceedings of the Physical Society}
}

@book{fetter,
author = {Fetter, A. L. and Walecka, J. D.},
title = {Quantum Theory of Many-Particle Systems},
publisher = {McGraw-Hill},
year = {1971},
}

@article{Danby1988,
  title = {Positron-Hf Collisions: Prediction of a Weakly Bound State},
  author = {Danby, Grahame and Tennyson, Jonathan},
  journal = {Phys. Rev. Lett.},
  volume = {61},
  issue = {24},
  pages = {2737--2739},
  numpages = {0},
  year = {1988},
  month = {12},
  publisher = {American Physical Society},
  doi = {10.1103/PhysRevLett.61.2737},
  url = {https://link.aps.org/doi/10.1103/PhysRevLett.61.2737}
}

@article{Dunning1992,
  title={Electron affinities of the first-row atoms revisited. Systematic basis sets and wave functions},
  author={Kendall, R. A. and Dunning Jr, T.H. and Harrison, R. J.},
  journal={J. Chem. Phys.},
  volume={96},
  pages={6796},
  year={1992},
}

@Inbook{Gribakin2001,
author="Gribakin, Gleb",
editor="Surko, Clifford M.
and Gianturco, Franco A.",
title="Theory of Positron Annihilation on Molecules",
bookTitle="New Directions in Antimatter Chemistry and Physics",
year="2001",
publisher="Springer Netherlands",
address="Dordrecht",
pages="413",
isbn="978-0-306-47613-6",
doi="10.1007/0-306-47613-4_22",
url="https://doi.org/10.1007/0-306-47613-4_22"
}

@article{Bubin2004,
    author = {Bubin, Sergiy and Adamowicz, Ludwik},
    title = {Non-Born–Oppenheimer study of positronic molecular systems: e+LiH},
    journal = {J.~Chem.~Phys.},
    volume = {120},
    number = {13},
    pages = {6051-6055},
    year = {2004},
    month = {04},
    issn = {0021-9606},
    doi = {10.1063/1.1651056},
    url = {https://doi.org/10.1063/1.1651056},
    eprint = {https://pubs.aip.org/aip/jcp/article-pdf/120/13/6051/19267046/6051_1_online.pdf},
}

@article{Gribakin2004,
  title = {Many-body theory of positron-atom interactions},
  author = {Gribakin, G. F. and Ludlow, J.},
  journal = {Phys. Rev. A},
  volume = {70},
  issue = {3},
  pages = {032720},
  numpages = {19},
  year = {2004},
  month = {9},
  publisher = {American Physical Society},
  doi = {10.1103/PhysRevA.70.032720},
  url = {https://link.aps.org/doi/10.1103/PhysRevA.70.032720}
}

@article{Gianturco2006,
  title = {Positron binding to alkali-metal hydrides: The role of molecular vibrations},
  author = {Gianturco, Franco A. and Franz, Jan and Buenker, Robert J. and Liebermann, Heinz-Peter and Pichl, Luk\'a\ifmmode \check{s}\else \v{s}\fi{} and Rost, Jan-Michael and Tachikawa, Masanori and Kimura, Mineo},
  journal = {Phys. Rev. A},
  volume = {73},
  issue = {2},
  pages = {022705},
  numpages = {9},
  year = {2006},
  month = {2},
  publisher = {American Physical Society},
  doi = {10.1103/PhysRevA.73.022705},
  url = {https://link.aps.org/doi/10.1103/PhysRevA.73.022705}
}

@article{Buenker2008,
title = {Configuration interaction calculations of positron binding to molecular oxides and hydrides and its effect on spectroscopic constants},
journal = {Nuclear Instruments and Methods in Physics Research Section B: Beam Interactions with Materials and Atoms},
volume = {266},
number = {3},
pages = {483},
year = {2008},
note = {Low Energy Positron and Positronium Physics},
issn = {0168-583X},
doi = {https://doi.org/10.1016/j.nimb.2007.12.029},
url = {https://www.sciencedirect.com/science/article/pii/S0168583X07017995},
author = {Robert J. Buenker and Heinz-Peter Liebermann}
}

@article{Danielson09,
author={J R Danielson and J A Young and C M Surko},
title={Dependence of positron-molecule binding energies on molecular properties},
journal={J. Phys. B},
volume={42},
pages={235203},
year={2009},
doi={10.1088/0953-4075/42/23/235203}
}

@article{Kita2009,
    author = {Kita, Yukiumi and Maezono, Ryo and Tachikawa, Masanori and Towler, Mike and Needs, Richard J.},
    title = {Ab initio quantum Monte Carlo study of the positronic hydrogen cyanide molecule},
    journal = {J.~Chem.~Phys.},
    volume = {131},
    number = {13},
    pages = {134310},
    year = {2009},
    month = {10},
    issn = {0021-9606},
    doi = {10.1063/1.3239502},
    url = {https://doi.org/10.1063/1.3239502},
    eprint = {https://pubs.aip.org/aip/jcp/article-pdf/doi/10.1063/1.3239502/14708949/134310_1_online.pdf},
}

@article{cassidy2018,
author={Cassidy, David B.},
year={2018},
title={Experimental progress in positronium laser physics},
journal={Eur. Phys. J. D},
pages={53},
volume={72},
doi={10.1140/epjd/e2018-80721-y}
}

@Article{alpha2,
title = {Confinement of antihydrogen for 1,000 seconds},
author = {{The ALPHA collaboration}},
Journal= {Nat Phys},
volume = {7},
pages = {558-564},
year = {2011},
publisher = {Nature Publishing Group, a division of Macmillan Publishers Limited.},
doi = {10.1038/nphys2025}
}

@Article{alpha1,
title = {Trapped antihydrogen},
author = {{The ALPHA collaboration}},
Journal= {Nat Phys},
volume = {468},
pages = {673-676},
year = {2010},
publisher = {Nature Publishing Group, a division of Macmillan Publishers Limited.},
doi = {10.1038/nature09610}
}

@article{ATRAP:2012,
  title = {Trapped Antihydrogen in Its Ground State},
  author = {Gabrielse, G. and Kalra, R. and Kolthammer, W. S. and McConnell, R. and Richerme, P. and Grzonka, D. and Oelert, W. and Sefzick, T. and Zielinski, M. and Fitzakerley, D. W. and George, M. C. and Hessels, E. A. and Storry, C. H. and Weel, M. and M\"ullers, A. and Walz, J.},
  collaboration = {ATRAP Collaboration},
  journal = {Phys. Rev. Lett.},
  volume = {108},
  issue = {11},
  pages = {113002},
  numpages = {4},
  year = {2012},
  month = {3},
  publisher = {American Physical Society},
  doi = {10.1103/PhysRevLett.108.113002},
}

@Article{Amole:2012,
author={Amole, C.
and Ashkezari, M. D.
and Baquero-Ruiz, M.
and Bertsche, W.
and Bowe, P. D.
and Butler, E.
and Capra, A.
and Cesar, C. L.
and Charlton, M.
and Deller, A.
and Donnan, P. H.
and Eriksson, S.
and Fajans, J.
and Friesen, T.
and Fujiwara, M. C.
and Gill, D. R.
and Gutierrez, A.
and Hangst, J. S.
and Hardy, W. N.
and Hayden, M. E.
and Humphries, A. J.
and Isaac, C. A.
and Jonsell, S.
and Kurchaninov, L.
and Little, A.
and Madsen, N.
and McKenna, J. T. K.
and Menary, S.
and Napoli, S. C.
and Nolan, P.
and Olchanski, K.
and Olin, A.
and Pusa, P.
and Rasmussen, C. {\O}
and Robicheaux, F.
and Sarid, E.
and Shields, C. R.
and Silveira, D. M.
and Stracka, S.
and So, C.
and Thompson, R. I.
and van der Werf, D. P.
and Wurtele, J. S.},
title={Resonant quantum transitions in trapped antihydrogen atoms},
journal={Nature},
year={2012},
month={3},
day={01},
volume={483},
number={7390},
pages={439-443},
issn={1476-4687},
doi={10.1038/nature10942},
url={https://doi.org/10.1038/nature10942}
}

@Article{Amole:2014,
author={Amole, C.
and Ashkezari, M. D.
and Baquero-Ruiz, M.
and Bertsche, W.
and Butler, E.
and Capra, A.
and Cesar, C. L.
and Charlton, M.
and Eriksson, S.
and Fajans, J.
and Friesen, T.
and Fujiwara, M. C.
and Gill, D. R.
and Gutierrez, A.
and Hangst, J. S.
and Hardy, W. N.
and Hayden, M. E.
and Isaac, C. A.
and Jonsell, S.
and Kurchaninov, L.
and Little, A.
and Madsen, N.
and McKenna, J. T. K.
and Menary, S.
and Napoli, S. C.
and Nolan, P.
and Olchanski, K.
and Olin, A.
and Povilus, A.
and Pusa, P.
and Rasmussen, C. {\O}
and Robicheaux, F.
and Sarid, E.
and Silveira, D. M.
and So, C.
and Tharp, T. D.
and Thompson, R. I.
and van der Werf, D. P.
and Vendeiro, Z.
and Wurtele, J. S.
and Zhmoginov, A. I.
and Charman, A. E.},
title={An experimental limit on the charge of antihydrogen},
journal={Nat.~Commun.},
year={2014},
month={6},
day={03},
volume={5},
number={1},
pages={3955},
issn={2041-1723},
doi={10.1038/ncomms4955},
url={https://doi.org/10.1038/ncomms4955}
}

@Article{GBAR,
author={Adrich, P.
and Blumer, P.
and Caratsch, G.
and Chung, M.
and Clad{\'e}, P.
and Comini, P.
and Crivelli, P.
and Dalkarov, O.
and Debu, P.
and Douillet, A.
and Drapier, D.
and Froelich, P.
and Garroum, N.
and Guellati-Khelifa, S.
and Guyomard, J.
and Hervieux, P.-A.
and Hilico, L.
and Indelicato, P.
and Jonsell, S.
and Karr, J.-P.
and Kim, B.
and Kim, S.
and Kim, E.-S.
and Ko, Y. J.
and Kosinski, T.
and Kuroda, N.
and Latacz, B. M.
and Lee, B.
and Lee, H.
and Lee, J.
and Lim, E.
and Liszkay, L.
and Lunney, D.
and Manfredi, G.
and Mansouli{\'e}, B.
and Matusiak, M.
and Nesvizhevsky, V.
and Nez, F.
and Niang, S.
and Ohayon, B.
and Park, K.
and Paul, N.
and P{\'e}rez, P.
and Regenfus, C.
and Reynaud, S.
and Roumegou, C.
and Rouss{\'e}, J.-Y.
and Sacquin, Y.
and Sadowski, G.
and Sarkisyan, J.
and Sato, M.
and Schmidt-Kaler, F.
and Staszczak, M.
and Szymczyk, K.
and Tanaka, T. A.
and Tuchming, B.
and Vallage, B.
and Voronin, A.
and van der Werf, D. P.
and Welker, A.
and Won, D.
and Wronka, S.
and Yamazaki, Y.
and Yoo, K.-H.
and Yzombard, P.},
title={Production of antihydrogen atoms by 6 keV antiprotons through a positronium cloud},
journal={Eur.~Phys.~J. C},
year={2023},
month={Nov},
day={06},
volume={83},
number={11},
pages={1004},
issn={1434-6052},
doi={10.1140/epjc/s10052-023-12137-y},
url={https://doi.org/10.1140/epjc/s10052-023-12137-y}
}

@Article{Ahmadi:2017,
author={Ahmadi, M.
and Alves, B. X. R.
and Baker, C. J.
and Bertsche, W.
and Butler, E.
and Capra, A.
and Carruth, C.
and Cesar, C. L.
and Charlton, M.
and Cohen, S.
and Collister, R.
and Eriksson, S.
and Evans, A.
and Evetts, N.
and Fajans, J.
and Friesen, T.
and Fujiwara, M. C.
and Gill, D. R.
and Gutierrez, A.
and Hangst, J. S.
and Hardy, W. N.
and Hayden, M. E.
and Isaac, C. A.
and Ishida, A.
and Johnson, M. A.
and Jones, S. A.
and Jonsell, S.
and Kurchaninov, L.
and Madsen, N.
and Mathers, M.
and Maxwell, D.
and McKenna, J. T. K.
and Menary, S.
and Michan, J. M.
and Momose, T.
and Munich, J. J.
and Nolan, P.
and Olchanski, K.
and Olin, A.
and Pusa, P.
and Rasmussen, C. {\O}
and Robicheaux, F.
and Sacramento, R. L.
and Sameed, M.
and Sarid, E.
and Silveira, D. M.
and Stracka, S.
and Stutter, G.
and So, C.
and Tharp, T. D.
and Thompson, J. E.
and Thompson, R. I.
and van der Werf, D. P.
and Wurtele, J. S.},
title={Antihydrogen accumulation for fundamental symmetry tests},
journal={Nat.~Commun.},
year={2017},
month={9},
day={25},
volume={8},
number={1},
pages={681},
issn={2041-1723},
doi={10.1038/s41467-017-00760-9},
url={https://doi.org/10.1038/s41467-017-00760-9}
}

@Article{Ahmadi:2016,
author={Ahmadi, M.
and Baquero-Ruiz, M.
and Bertsche, W.
and Butler, E.
and Capra, A.
and Carruth, C.
and Cesar, C. L.
and Charlton, M.
and Charman, A. E.
and Eriksson, S.
and Evans, L. T.
and Evetts, N.
and Fajans, J.
and Friesen, T.
and Fujiwara, M. C.
and Gill, D. R.
and Gutierrez, A.
and Hangst, J. S.
and Hardy, W. N.
and Hayden, M. E.
and Isaac, C. A.
and Ishida, A.
and Jones, S. A.
and Jonsell, S.
and Kurchaninov, L.
and Madsen, N.
and Maxwell, D.
and McKenna, J. T. K.
and Menary, S.
and Michan, J. M.
and Momose, T.
and Munich, J. J.
and Nolan, P.
and Olchanski, K.
and Olin, A.
and Povilus, A.
and Pusa, P.
and Rasmussen, C. {\O}
and Robicheaux, F.
and Sacramento, R. L.
and Sameed, M.
and Sarid, E.
and Silveira, D. M.
and So, C.
and Tharp, T. D.
and Thompson, R. I.
and van der Werf, D. P.
and Wurtele, J. S.
and Zhmoginov, A. I.},
title={An improved limit on the charge of antihydrogen from stochastic acceleration},
journal={Nature},
year={2016},
month={1},
day={01},
volume={529},
number={7586},
pages={373-376},
issn={1476-4687},
doi={10.1038/nature16491},
url={https://doi.org/10.1038/nature16491}
}

@article{Malbrunot18,
author = {Malbrunot, C.  and Amsler, C.  and Arguedas Cuendis, S.  and Breuker, H.  and Dupre, P.  and Fleck, M.  and Higaki, H.  and Kanai, Y.  and Kolbinger, B.  and Kuroda, N.  and Leali, M.  and M{\"a}ckel, V.  and Mascagna, V.  and Massiczek, O.  and Matsuda, Y.  and Nagata, Y.  and Simon, M. C.  and Spitzer, H.  and Tajima, M.  and Ulmer, S.  and Venturelli, L.  and Widmann, E.  and Wiesinger, M.  and Yamazaki, Y.  and Zmeskal, J. },
title = {The ASACUSA antihydrogen and hydrogen program: results and prospects},
journal = {Philos. Trans. Roy. Soc. A},
volume = {376},
number = {2116},
pages = {20170273},
year = {2018},
doi = {10.1098/rsta.2017.0273}
}

@Article{Baker2021,
author={Baker, C. J.
and Bertsche, W.
and Capra, A.
and Carruth, C.
and Cesar, C. L.
and Charlton, M.
and Christensen, A.
and Collister, R.
and Mathad, A. Cridland
and Eriksson, S.
and Evans, A.
and Evetts, N.
and Fajans, J.
and Friesen, T.
and Fujiwara, M. C.
and Gill, D. R.
and Grandemange, P.
and Granum, P.
and Hangst, J. S.
and Hardy, W. N.
and Hayden, M. E.
and Hodgkinson, D.
and Hunter, E.
and Isaac, C. A.
and Johnson, M. A.
and Jones, J. M.
and Jones, S. A.
and Jonsell, S.
and Khramov, A.
and Knapp, P.
and Kurchaninov, L.
and Madsen, N.
and Maxwell, D.
and McKenna, J. T. K.
and Menary, S.
and Michan, J. M.
and Momose, T.
and Mullan, P. S.
and Munich, J. J.
and Olchanski, K.
and Olin, A.
and Peszka, J.
and Powell, A.
and Pusa, P.
and Rasmussen, C. {\O}
and Robicheaux, F.
and Sacramento, R. L.
and Sameed, M.
and Sarid, E.
and Silveira, D. M.
and Starko, D. M.
and So, C.
and Stutter, G.
and Tharp, T. D.
and Thibeault, A.
and Thompson, R. I.
and van der Werf, D. P.
and Wurtele, J. S.},
title={Laser cooling of antihydrogen atoms},
journal={Nature},
year={2021},
month={4},
day={01},
volume={592},
number={7852},
pages={35},
issn={1476-4687},
doi={10.1038/s41586-021-03289-6},
url={https://doi.org/10.1038/s41586-021-03289-6}
}

@article{Amsler21,
author={Claude Amsler and Massimiliano Antonello and Alexander Belov and Germano Bonomi and Roberto Sennen Brusa and Massimo Caccia and Antoine Camper and Ruggero Caravita and Fabrizio Castelli and Patrick Cheinet and Daniel Comparat and Giovanni Consolati and Andrea Demetrio and Lea Di Noto and Michael Doser and Mattia Fan{\`i} and Rafael Ferragut and Julian Fesel and Sebastian Gerber and Marco Giammarchi and Angela Gligorova and Lisa Theresa Gl{\"o}ggler and Francesco Guatieri and Stefan Haider and Alexander Hinterberger and Alban Kellerbauer and Olga Khalidova and Daniel Krasnicky and Vittorio Lagomarsino and Chlo{\'e} Malbrunot and Sebastiano Mariazzi and Viktor Matveev and Simon Muller and Giancarlo Nebbia and Patrick Nedelec and Lilian Nowak and Markus Oberthaler and Emmanuel Oswald and Davide Pagano and Luca Penasa and Vojtech Petracek and Luca Povolo and Francesco Prelz and Marco Prevedelli and Benjamin Rien{\"a}cker and Ole R{\o}hne and Alberto Rotondi and Heidi Sandaker and Romualdo Santoro and Gemma Testera and Ingmari Tietje and Valerio Toso and Tim Wolz and Pauline Yzombard and Christian Zimmer and Nicola Zurlo},
title={Pulsed production of antihydrogen},
journal={Comm. Phys.},
volume={4},
pages={19},
year={2021},
doi={10.1038/s42005-020-00494-z}
}

@Article{Adrich2023,
author={Adrich, P.
and Blumer, P.
and Caratsch, G.
and Chung, M.
and Clad{\'e}, P.
and Comini, P.
and Crivelli, P.
and Dalkarov, O.
and Debu, P.
and Douillet, A.
and Drapier, D.
and Froelich, P.
and Garroum, N.
and Guellati-Khelifa, S.
and Guyomard, J.
and Hervieux, P.-A.
and Hilico, L.
and Indelicato, P.
and Jonsell, S.
and Karr, J.-P.
and Kim, B.
and Kim, S.
and Kim, E.-S.
and Ko, Y. J.
and Kosinski, T.
and Kuroda, N.
and Latacz, B. M.
and Lee, B.
and Lee, H.
and Lee, J.
and Lim, E.
and Liszkay, L.
and Lunney, D.
and Manfredi, G.
and Mansouli{\'e}, B.
and Matusiak, M.
and Nesvizhevsky, V.
and Nez, F.
and Niang, S.
and Ohayon, B.
and Park, K.
and Paul, N.
and P{\'e}rez, P.
and Regenfus, C.
and Reynaud, S.
and Roumegou, C.
and Rouss{\'e}, J.-Y.
and Sacquin, Y.
and Sadowski, G.
and Sarkisyan, J.
and Sato, M.
and Schmidt-Kaler, F.
and Staszczak, M.
and Szymczyk, K.
and Tanaka, T. A.
and Tuchming, B.
and Vallage, B.
and Voronin, A.
and van der Werf, D. P.
and Welker, A.
and Won, D.
and Wronka, S.
and Yamazaki, Y.
and Yoo, K.-H.
and Yzombard, P.},
title={Production of antihydrogen atoms by 6 keV antiprotons through a positronium cloud},
journal={Eur.~Phys.~J. C},
year={2023},
month={11},
day={06},
volume={83},
number={11},
pages={1004},
issn={1434-6052},
doi={10.1140/epjc/s10052-023-12137-y},
url={https://doi.org/10.1140/epjc/s10052-023-12137-y}
}

@Article{Anderson2023,
author={Anderson, E. K.
and Baker, C. J.
and Bertsche, W.
and Bhatt, N. M.
and Bonomi, G.
and Capra, A.
and Carli, I.
and Cesar, C. L.
and Charlton, M.
and Christensen, A.
and Collister, R.
and Cridland Mathad, A.
and Duque Quiceno, D.
and Eriksson, S.
and Evans, A.
and Evetts, N.
and Fabbri, S.
and Fajans, J.
and Ferwerda, A.
and Friesen, T.
and Fujiwara, M. C.
and Gill, D. R.
and Golino, L. M.
and Gomes Gon{\c{c}}alves, M. B.
and Grandemange, P.
and Granum, P.
and Hangst, J. S.
and Hayden, M. E.
and Hodgkinson, D.
and Hunter, E. D.
and Isaac, C. A.
and Jimenez, A. J. U.
and Johnson, M. A.
and Jones, J. M.
and Jones, S. A.
and Jonsell, S.
and Khramov, A.
and Madsen, N.
and Martin, L.
and Massacret, N.
and Maxwell, D.
and McKenna, J. T. K.
and Menary, S.
and Momose, T.
and Mostamand, M.
and Mullan, P. S.
and Nauta, J.
and Olchanski, K.
and Oliveira, A. N.
and Peszka, J.
and Powell, A.
and Rasmussen, C. {\O}
and Robicheaux, F.
and Sacramento, R. L.
and Sameed, M.
and Sarid, E.
and Schoonwater, J.
and Silveira, D. M.
and Singh, J.
and Smith, G.
and So, C.
and Stracka, S.
and Stutter, G.
and Tharp, T. D.
and Thompson, K. A.
and Thompson, R. I.
and Thorpe-Woods, E.
and Torkzaban, C.
and Urioni, M.
and Woosaree, P.
and Wurtele, J. S.},
title={Observation of the effect of gravity on the motion of antimatter},
journal={Nature},
year={2023},
month={9},
day={01},
volume={621},
number={7980},
pages={716-722},
issn={1476-4687},
doi={10.1038/s41586-023-06527-1},
url={https://doi.org/10.1038/s41586-023-06527-1}}

@article{Moskal2021,
	author = {Moskal, P. and Gajos, A. and Mohammed, M. and Chhokar, J. and Chug, N. and Curceanu, C. and Czerwi{\'n}ski, E. and Dadgar, M. and Dulski, K. and Gorgol, M. and Goworek, J. and Hiesmayr, B. C. and Jasi{\'n}ska, B. and Kacprzak, K. and Kap{\l}on, {\L}. and Karimi, H. and Kisielewska, D. and Klimaszewski, K. and Korcyl, G. and Kowalski, P. and Krawczyk, N. and Krzemie{\'n}, W. and Kozik, T. and Kubicz, E. and Nied{\'z}wiecki, S. and Parzych, S. and Pawlik-Nied{\'z}wiecka, M. and Raczy{\'n}ski, L. and Raj, J. and Sharma, S. and Choudhary, S. and Shopa, R. Y. and Sienkiewicz, A. and Silarski, M. and Skurzok, M. and St{\k{e}}pie{\'n}, E. {\L}. and Tayefi, F. and Wi{\'s}licki, W.},
	doi = {10.1038/s41467-021-25905-9},
	journal = {Nat. Commun.},
	number = {1},
	pages = {5658},
	title = {Testing CPT symmetry in ortho-positronium decays with positronium annihilation tomography},
	url = {https://doi.org/10.1038/s41467-021-25905-9},
	volume = {12},
	year = {2021}}

@article{Danielson:2015,
  title = {Plasma and trap-based techniques for science with positrons},
  author = {Danielson, J. R. and Dubin, D. H. E. and Greaves, R. G. and Surko, C. M.},
  journal = {Rev. Mod. Phys.},
  volume = {87},
  issue = {1},
  pages = {247},
  numpages = {60},
  year = {2015},
  month = {3},
  publisher = {American Physical Society},
  doi = {10.1103/RevModPhys.87.247}}

@article{RevModPhys.82.2557,
  title = {Positron-molecule interactions: Resonant attachment, annihilation, and bound states},
  author = {Gribakin, G. F. and Young, J. A. and Surko, C. M.},
  journal = {Rev. Mod. Phys.},
  volume = {82},
  issue = {3},
  pages = {2557},
  numpages = {0},
  year = {2010},
  month = {9},
  publisher = {American Physical Society},
  doi = {10.1103/RevModPhys.82.2557},
  url = {https://link.aps.org/doi/10.1103/RevModPhys.82.2557}
}

@article{Prantzos2011,
  title = {The 511 keV emission from positron annihilation in the Galaxy},
  author = {Prantzos, N. and Boehm, C. and Bykov, A. M. and Diehl, R. and Ferri\`ere, K. and Guessoum, N. and Jean, P. and Knoedlseder, J. and Marcowith, A. and Moskalenko, I. V. and Strong, A. and Weidenspointner, G.},
  journal = {Rev. Mod. Phys.},
  volume = {83},
  issue = {3},
  pages = {1001},
  numpages = {0},
  year = {2011},
  month = {9},
  publisher = {American Physical Society},
  doi = {10.1103/RevModPhys.83.1001},
  url = {https://link.aps.org/doi/10.1103/RevModPhys.83.1001}
}

@Article{PhysRevA.52.4541,
  title = {Bound states of positrons and neutral atoms},
  author = {Dzuba, V. A. and Flambaum, V. V. and Gribakin, G. F. and King, W. A.},
  journal = {Phys. Rev. A},
  volume = {52},
  number = {6},
  pages = {4541},
  numpages = {5},
  year = {1995},
  month = {12},
  doi = {10.1103/PhysRevA.52.4541},
  publisher = {American Physical Society}
}

@article{Amusia:Pos:MBT:He,
  author={M. {Ya} Amusia and N. A. Cherepkov and L. V. Chernysheva and S. G. Shapiro},
  title={Elastic scattering of slow positrons by helium},
  journal={J. Phys. B: Atom. Mol. Phys.},
  volume={9},
  number={17},
  pages={L531},
  url={http://stacks.iop.org/0022-3700/9/i=17/a=005},
  year={1976}
}

@article{Cederbaum-elecpos,
  title = {Many-body theory of composite electronic-positronic systems},
  author = {M\"uller, M. and Cederbaum, L. S.},
  journal = {Phys. Rev. A},
  volume = {42},
  issue = {1},
  pages = {170--183},
  numpages = {0},
  year = {1990},
  month = {7},
  publisher = {American Physical Society},
  doi = {10.1103/PhysRevA.42.170},
  url = {https://link.aps.org/doi/10.1103/PhysRevA.42.170}
}

@article{PhysScripta.46.248,
  author={V. A. Dzuba and V. V. Flambaum and W. A. King and B. N. Miller and O. P. Sushkov},
  title={Interaction between slow positrons and atoms},
  journal={Phys. Scr.},
  volume={T46},
  number={T46},
  pages={248},
  doi={10.1088/0031-8949/1993/T46/039},
  url={http://stacks.iop.org/1402-4896/1993/i=T46/a=039},
  year={1993}}

@Article{Gribakin:2004,
  title = {Many-body theory of positron-atom interactions},
  author = {Gribakin, G. F. and Ludlow, J. },
  journal = {Phys. Rev. A},
  volume = {70},
  number = {3},
  pages = {032720},
  numpages = {19},
  year = {2004},
  month = {9},
  doi = {10.1103/PhysRevA.70.032720},
  publisher = {American Physical Society}}

@article{Flambaum:2021,
  title = {Radiation from matter-antimatter annihilation in the quark nugget model of dark matter},
  author = {Flambaum, V. V. and Samsonov, I. B.},
  journal = {Phys. Rev. D},
  volume = {104},
  issue = {6},
  pages = {063042},
  numpages = {17},
  year = {2021},
  month = {9},
  publisher = {American Physical Society},
  doi = {10.1103/PhysRevD.104.063042},
}

@article{posfun,
    author = {Drachman, Richard J.},
    title = "{Why positron physics is fun}",
    journal = {AIP Conference Proceedings},
    volume = {360},
    number = {1},
    pages = {369},
    year = {1996},
    month = {02},
    issn = {0094-243X},
    doi = {10.1063/1.49828},
}

@article{Fuller:2019,
  title = {Positrons and 511 keV Radiation as Tracers of Recent Binary Neutron Star Mergers},
  author = {Fuller, George M. and Kusenko, Alexander and Radice, David and Takhistov, Volodymyr},
  journal = {Phys. Rev. Lett.},
  volume = {122},
  issue = {12},
  pages = {121101},
  numpages = {5},
  year = {2019},
  month = {3},
  publisher = {American Physical Society},
  doi = {10.1103/PhysRevLett.122.121101},
}

@article{harabati2014identification,
  title={Identification of atoms that can bind positrons},
  author={Harabati, C and Dzuba, VA and Flambaum, VV},
  journal={Phys. Rev. A.},
  volume={89},
  number={2},
  pages={022517},
  year={2014},
  publisher={APS}
}

@article{Tachikawa2011,
author ={Tachikawa, Masanori and Kita, Yukiumi and Buenker, Robert J.},
title  ={Bound states of the positron with nitrile species with a configuration interaction multi-component molecular orbital approach},
journal  ={Phys. Chem. Chem. Phys.},
year  ={2011},
volume  ={13},
issue  ={7},
pages  ={2701-2705},
publisher  ={The Royal Society of Chemistry},
doi  ={10.1039/C0CP01650K},
url  ={http://dx.doi.org/10.1039/C0CP01650K}
}

@article{Koyanagi2012,
    title={Systematic theoretical investigation of a positron binding to amino acid molecules using the ab initio multi-component molecular orbital approach.},
    author={Koyanagi, K. and Kita, Y. and Tachikawa, M.},
    journal={Eur. Phys. J. D},
    volume={66},
    pages={121},
    year={2012}
}

@article{Tuomisto2013,
author = {F Tuomisto and I Makkonen},
journal = {Rev. Mod. Phys.},
title = {Defect identification in semiconductors with positron annihilation: Experiment and theory},
volume = {85},
pages = {1583},
year = {2013}}

@article{Green2013,
  title = {Positron scattering and annihilation in hydrogenlike ions},
  author = {Green, D. G. and Gribakin, G. F.},
  journal = {Phys. Rev. A},
  volume = {88},
  issue = {3},
  pages = {032708},
  numpages = {24},
  year = {2013},
  month = {9},
  publisher = {American Physical Society},
  doi = {10.1103/PhysRevA.88.032708},
  url = {https://link.aps.org/doi/10.1103/PhysRevA.88.032708}
}

@article{Charry2014,
  title = {Calculation of positron binding energies of amino acids with the any-particle molecular-orbital approach},
  author = {Charry, J. and Romero, J. and Varella, M. T. do N. and Reyes, A.},
  journal = {Phys. Rev. A},
  volume = {89},
  issue = {5},
  pages = {052709},
  numpages = {11},
  year = {2014},
  month = {5},
  publisher = {American Physical Society},
  doi = {10.1103/PhysRevA.89.052709},
  url = {https://link.aps.org/doi/10.1103/PhysRevA.89.052709}
}

@article{Romero2014,
    author = {Romero, Jonathan and Charry, Jorge A. and Flores-Moreno, Roberto and Varella, Márcio T. do N. and Reyes, Andrés},
    title = {Calculation of positron binding energies using the generalized any particle propagator theory},
    journal = {J.~Chem.~Phys.},
    volume = {141},
    number = {11},
    pages = {114103},
    year = {2014},
    month = {09},
    issn = {0021-9606},
    doi = {10.1063/1.4895043},
    url = {https://doi.org/10.1063/1.4895043},
    eprint = {https://pubs.aip.org/aip/jcp/article-pdf/doi/10.1063/1.4895043/15484085/114103_1_online.pdf},
}

@article{Green2014,
  title = {Positron scattering and annihilation on noble-gas atoms},
  author = {Green, D. G. and Ludlow, J. A. and Gribakin, G. F.},
  journal = {Phys. Rev. A},
  volume = {90},
  issue = {3},
  pages = {032712},
  numpages = {19},
  year = {2014},
  month = {9},
  publisher = {American Physical Society},
  doi = {10.1103/PhysRevA.90.032712},
  url = {https://link.aps.org/doi/10.1103/PhysRevA.90.032712}
}

@article{Tachikawa2014,
doi = {10.1088/1742-6596/488/1/012053},
url = {https://doi.org/10.1088/1742-6596/488/1/012053},
year = {2014},
publisher = {},
volume = {488},
number = {1},
pages = {012053},
author = {Tachikawa, Masanori},
title = {Positron-attachment to acetonitrile, acetaldehyde, and acetone molecules: Vibrational enhancement of positron affinities with configuration interaction level of multi-component molecular orbital approach},
journal = {J.~Phys.: Conference Series}
}

@article{Yamada2014,
  title = {Theoretical prediction of the binding of a positron to a formaldehyde molecule using a first-principles calculation},
  author = {Yamada, Yurika and Kita, Yukiumi and Tachikawa, Masanori},
  journal = {Phys. Rev. A},
  volume = {89},
  issue = {6},
  pages = {062711},
  numpages = {5},
  year = {2014},
  month = {6},
  publisher = {American Physical Society},
  doi = {10.1103/PhysRevA.89.062711},
  url = {https://link.aps.org/doi/10.1103/PhysRevA.89.062711}
}

@article{Green2015,
  title = {$\ensuremath{\gamma}$ Spectra and Enhancement Factors for Positron Annihilation with Core Electrons },
  author = {Green, D. G. and Gribakin, G. F.},
  journal = {Phys. Rev. Lett.},
  volume = {114},
  issue = {9},
  pages = {093201},
  numpages = {5},
  year = {2015},
  month = {3},
  publisher = {American Physical Society},
  doi = {10.1103/PhysRevLett.114.093201},
  url =  {http://link.aps.org/doi/10.1103/PhysRevLett.114.093201}
}

@article{Hugenschmidt2016,
author = {C Hugenschmidt},
journal = {Surf. Sci. Rep.},
title = {Positrons in surface physics},
volume = {71},
pages = {547},
year = {2016}}

@article{Gribakin2017,
  title = {Mode coupling and multiquantum vibrational excitations in Feshbach-resonant positron annihilation in molecules},
  author = {Gribakin, G. F. and Stanton, J. F. and Danielson, J. R. and Natisin, M. R. and Surko, C. M.},
  journal = {Phys. Rev. A},
  volume = {96},
  issue = {6},
  pages = {062709},
  numpages = {16},
  year = {2017},
  month = {12},
  publisher = {American Physical Society},
  doi = {10.1103/PhysRevA.96.062709},
  url = {https://link.aps.org/doi/10.1103/PhysRevA.96.062709}
}

@article{Moreira2019,
author = {Moreira, Giseli M. and Bettega, Márcio H. F.},
title = {Elastic Scattering of Slow Positrons by Pyrazine},
journal = {J.~Phys.~Chem. A},
volume = {123},
number = {42},
pages = {9132},
year = {2019},
doi = {10.1021/acs.jpca.9b08845},
URL = { 
    https://doi.org/10.1021/acs.jpca.9b08845
},
eprint = {       https://doi.org/10.1021/acs.jpca.9b08845
}
}

@article{Barbosa2019,
  title = {Theoretical study on positron scattering by benzene over a broad energy range},
  author = {Barbosa, Alessandra Souza and Blanco, Francisco and Garc\'{\i}a, Gustavo and Bettega, M. H. F.},
  journal = {Phys. Rev. A},
  volume = {100},
  issue = {4},
  pages = {042705},
  numpages = {7},
  year = {2019},
  month = {10},
  publisher = {American Physical Society},
  doi = {10.1103/PhysRevA.100.042705},
  url = {https://link.aps.org/doi/10.1103/PhysRevA.100.042705}
}

@article{Swann2019,
  title = {Positron Binding and Annihilation in Alkane Molecules},
  author = {Swann, A. R. and Gribakin, G. F.},
  journal = {Phys. Rev. Lett.},
  volume = {123},
  issue = {11},
  pages = {113402},
  numpages = {6},
  year = {2019},
  month = {9},
  publisher = {American Physical Society},
  doi = {10.1103/PhysRevLett.123.113402},
}

@article{Swann2020,
    author = {Swann, A. R. and Gribakin, G. F.},
    title = {Effect of molecular constitution and conformation on positron binding and annihilation in alkanes},
    journal = {J.~Chem.~Phys.},
    volume = {153},
    number = {18},
    pages = {184311},
    year = {2020},
    month = {11},
    issn = {0021-9606},
    doi = {10.1063/5.0028071},
    url = {https://doi.org/10.1063/5.0028071},
    eprint = {https://pubs.aip.org/aip/jcp/article-pdf/doi/10.1063/5.0028071/15581535/184311\_1\_online.pdf},
}

@article{Edwards2021,
  title = {Positron scattering from pyrazine},
  author = {Edwards, D. and Stevens, D. and Cheong, Z. and Graves, V. and Gorfinkiel, J. D. and Blanco, F. and Garcia, G. and Brunger, M. J. and White, R. D. and Sullivan, J. P.},
  journal = {Phys. Rev. A},
  volume = {104},
  issue = {4},
  pages = {042807},
  numpages = {12},
  year = {2021},
  month = {10},
  publisher = {American Physical Society},
  doi = {10.1103/PhysRevA.104.042807},
  url = {https://link.aps.org/doi/10.1103/PhysRevA.104.042807}
}

@article{Swann2021,
  title = {Effect of chlorination on positron binding to hydrocarbons: Experiment and theory},
  author = {Swann, A. R. and Gribakin, G. F. and Danielson, J. R. and Ghosh, S. and Natisin, M. R. and Surko, C. M.},
  journal = {Phys. Rev. A},
  volume = {104},
  issue = {1},
  pages = {012813},
  numpages = {15},
  year = {2021},
  month = {7},
  publisher = {American Physical Society},
  doi = {10.1103/PhysRevA.104.012813},
  url = {https://link.aps.org/doi/10.1103/PhysRevA.104.012813}
}

@article{Ozaki2021,
author = {Ozaki, Maya and Yoshida, Daisuke and Kita, Yukiumi and Shimazaki, Tomomi and Tachikawa, Masanori},
title = {Positron Binding and Annihilation Properties of Amino Acid Systems},
journal = {ACS Omega},
volume = {6},
number = {44},
pages = {29449},
year = {2021},
doi = {10.1021/acsomega.1c03409},
URL = {  https://doi.org/10.1021/acsomega.1c03409}
}

@article{Graves2022,
  author       = {Graves, Vincent and Gorfinkiel, Jimena D.},
  title        = {R-matrix calculations for elastic electron and positron scattering from pyrazine: effect of the polarization description},
  journal = {Eur.~Phys.~J. D},
  date         = {2022-03-10},
  year         = {2022},
  volume       = {76},
  number       = {3},
  pages        = {43},
  doi          = {10.1140/epjd/s10053-022-00371-0},
  url          = {https://doi.org/10.1140/epjd/s10053-022-00371-0},
  issn         = {1434-6079}
}

@ARTICLE{Hofierka2022,
   author       = {J. Hofierka and B. Cunningham and C. M. Rawlins and C. H. Patterson and D. G. Green},
   title        = {Many-Body Theory of Positron Binding to Polyatomic Molecules},
   journal      = "Nature",
   volume       = "606",
   pages        = "688",
   year         = "2022",
}

@article{ArthurBaidoo2024,
  title = {Positron annihilation and binding in aromatic and other ring molecules},
  author = {Arthur-Baidoo, E. and Danielson, J. R. and Surko, C. M. and Cassidy, J. P. and Gregg, S. K. and Hofierka, J. and Cunningham, B. and Patterson, C. H. and Green, D. G.},
  journal = {Phys. Rev. A},
  volume = {109},
  issue = {6},
  pages = {062801},
  numpages = {14},
  year = {2024},
  month = {6},
  publisher = {American Physical Society},
  doi = {10.1103/PhysRevA.109.062801},
  url = {https://link.aps.org/doi/10.1103/PhysRevA.109.062801}
}

@article{Moreira2024,
    author={Moreira, G.M. and Bettega, M.H.F.},
    title={Can a positron bind to the para-benzoquinone molecule?},
    journal={Eur. Phys. J. D},
    volume={78},
    issue={9},
    year={2024},
    url={https://doi.org/10.1140/epjd/s10053-024-00800-2}
}

@article{Tachikawa:MCP_a,
  title = {Positron binding in chloroethenes: Modeling positron-electron correlation-polarization potentials for molecular calculations},
  author = {Suzuki, Haruya and Otomo, Takuma and Iida, Ryusei and Sugiura, Yutaro and Takayanagi, Toshiyuki and Tachikawa, Masanori},
  journal = {Phys. Rev. A},
  volume = {102},
  issue = {5},
  pages = {052830},
  numpages = {9},
  year = {2020},
  month = {11},
  publisher = {American Physical Society},
  doi = {10.1103/PhysRevA.102.052830},
  url = {https://link.aps.org/doi/10.1103/PhysRevA.102.052830}
}

@article{Swann:2020,
    author = {Swann, A. R. and Gribakin, G. F.},
    title = {Effect of molecular constitution and conformation on positron binding and annihilation in alkanes},
    journal = {J.~Chem.~Phys.},
    volume = {153},
    number = {18},
    pages = {184311},
    year = {2020},
    month = {11},
    issn = {0021-9606},
    doi = {10.1063/5.0028071},
    url = {https://doi.org/10.1063/5.0028071}
}

@ARTICLE{JPET2025,
  author={Moskal, Paweł and Bilewicz, Aleksander and Das, Manish and Huang, Bangyan and Khreptak, Aleksander and Parzych, Szymon and Qi, Jinyi and Rominger, Axel and Seifert, Robert and Sharma, Sushil and Shi, Kuangyu and Steinberger, William M. and Walczak, Rafał and Stępień, Ewa},
  journal={IEEE Transactions on Radiation and Plasma Medical Sciences}, 
  title={Positronium Imaging: History, Current Status, and Future Perspectives}, 
  year={2025},
  volume={9},
  number={8},
  pages={981},
  doi={10.1109/TRPMS.2025.3583554}}

@article{Moskal2024,
author = {Paweł Moskal  and Jakub Baran and others},
title = {Positronium image of the human brain in vivo},
journal = {Science Advances},
volume = {10},
number = {37},
pages = {eadp2840},
year = {2024},
doi = {10.1126/sciadv.adp2840},
URL = {https://www.science.org/doi/abs/10.1126/sciadv.adp2840},
eprint = {https://www.science.org/doi/pdf/10.1126/sciadv.adp2840}}

@article{Cassidy2024,
  title = {Many-body theory calculations of positron binding to halogenated hydrocarbons},
  author = {Cassidy, J. P. and Hofierka, J. and Cunningham, B. and Rawlins, C. M. and Patterson, C. H. and Green, D. G.},
  journal = {Phys. Rev. A},
  volume = {109},
  issue = {4},
  pages = {L040801},
  numpages = {6},
  year = {2024},
  month = {4},
  publisher = {American Physical Society},
  doi = {10.1103/PhysRevA.109.L040801},
  url = {https://link.aps.org/doi/10.1103/PhysRevA.109.L040801}
}

@article{Cassidy2024_2,
    author = {Cassidy, J. P. and Hofierka, J. and Cunningham, B. and Green, D. G.},
    title = {Many-body theory calculations of positronic-bonded molecular dianions},
    journal = {J.~Chem.~Phys.},
    volume = {160},
    number = {8},
    pages = {084304},
    year = {2024},
    month = {02},
    issn = {0021-9606},
    doi = {10.1063/5.0188719},
    url = {https://doi.org/10.1063/5.0188719},
    eprint = {https://pubs.aip.org/aip/jcp/article-pdf/doi/10.1063/5.0188719/19692847/084304\_1\_5.01887 19.pdf},
}

@ARTICLE{Hofierka2023,
author={Hofierka, J.  and Rawlins, C. M.  and Cunningham, B.  and Waide, D. T.  and Green, D. G. },   
title={Many-body theory calculations of positron scattering and annihilation in noble-gas atoms via the solution of Bethe–Salpeter equations using the Gaussian-basis code EXCITON+},        
journal={Front.~in Physics},        
volume={11},
year={2023}, 
number={2296-424X}
}

@article{BassRMP,
  title = {Colloquium: Positronium physics and biomedical applications},
  author = {Bass, Steven D. and Mariazzi, Sebastiano and Moskal, Pawel and {St\ifmmode \mbox{\k{e}}\else \k{e}\fi{}pie\ifmmode \acute{n}\else \'{n}\fi{}}, Ewa},
  journal = {Rev. Mod. Phys.},
  volume = {95},
  issue = {2},
  pages = {021002},
  numpages = {21},
  year = {2023},
  month = {5},
  publisher = {American Physical Society},
  doi = {10.1103/RevModPhys.95.021002},
  url = {https://link.aps.org/doi/10.1103/RevModPhys.95.021002}
}

@article{Rawlins2023,
  title = {Many-Body Theory Calculations of Positron Scattering and Annihilation in ${\mathrm{H}}_{2}$, ${\mathrm{N}}_{2}$, and ${\mathrm{CH}}_{4}$},
  author = {Rawlins, C. M. and Hofierka, J. and Cunningham, B. and Patterson, C. H. and Green, D. G.},
  journal = {Phys. Rev. Lett.},
  volume = {130},
  issue = {26},
  pages = {263001},
  numpages = {9},
  year = {2023},
  month = {6},
  publisher = {American Physical Society},
  doi = {10.1103/PhysRevLett.130.263001},
  url = {https://link.aps.org/doi/10.1103/PhysRevLett.130.263001}
}

@article{Hofierka2024,
  title = {Many-body theory calculations of positron binding to hydrogen cyanide},
  author = {Hofierka, J. and Cunningham, B. and Green, D. G.},
  journal = {Eur. Phys. J. D},
  volume = {78},
  pages = {37}, 
  year = {2024}
}

@article{Cassella2024,
  title = {Neural network variational Monte Carlo for positronic chemistry},
  author = {Cassella, G. and Foulkes, W. M. C. and Pfau, D. and Spencer, J. S.},
  journal = {Nat.~Commun.~},
  volume = {15},
  issue = {1}, 
  pages ={5214},
  year = {2024},
  doi = {10.1038/s41467-024-49290-1},
  url = {https://doi.org/10.1038/s41467-024-49290-1}
}

@article{Upadhyay2024,
author = {Upadhyay, Shiv and Benali, Anouar and Jordan, Kenneth D.},
title = {Capturing Correlation Effects in Positron Binding to Atoms and Molecules},
journal = {J.~Chem.~Theory Comput.},
volume = {20},
number = {22},
pages = {9879-9893},
year = {2024},
doi = {10.1021/acs.jctc.4c00727},
URL = { 
https://doi.org/10.1021/acs.jctc.4c00727
},
eprint = { https://doi.org/10.1021/acs.jctc.4c00727
}
}

@article{Ashiba2025,
author = {Ashiba, M. and Yoshida, D. and Kita, Y. and Shimazaki, T. and Takayanagi, T. and Tachikawa, M.},
title = {Effects of Halogenations and Conformational Isomers on Positron Binding in Halogenated Hydrocarbons.},
journal = {J. Comput. Chem.},
volume = {46},
number = {23},
pages = {e70217},
year = {2025},
doi = {10.1002/jcc.70217}
}

@article{Dzuba1994,
  title = {Correlation-potential method for negative ions and electron scattering},
  author = {Dzuba, V. A. and Gribakin, G. F.},
  journal = {Phys. Rev. A},
  volume = {49},
  issue = {4},
  pages = {2483},
  numpages = {0},
  year = {1994},
  month = {4},
  publisher = {American Physical Society},
  doi = {10.1103/PhysRevA.49.2483},
  url = {https://link.aps.org/doi/10.1103/PhysRevA.49.2483}
}

@article{Danielson22,
  title = {Enhancement of positron binding energy in molecules containing $\ensuremath{\pi}$ bonds},
  author = {Danielson, J. R. and Ghosh, S. and Surko, C. M.},
  journal = {Phys. Rev. A},
  volume = {106},
  issue = {3},
  pages = {032811},
  numpages = {8},
  year = {2022},
  month = {9},
  publisher = {American Physical Society},
  doi = {10.1103/PhysRevA.106.032811},
  url = {https://link.aps.org/doi/10.1103/PhysRevA.106.032811}
}

@article{Danielson25,
  title = {Improved positron-molecule binding energies and estimations using molecular parameters},
  author = {Danielson, J. R. and Arthur-Baidoo, E. and Surko, C. M.},
  journal = {Phys. Rev. A},
  volume = {111},
  issue = {4},
  pages = {042809},
  numpages = {16},
  year = {2025},
  month = {4},
  publisher = {American Physical Society},
  doi = {10.1103/PhysRevA.111.042809},
  url = {https://link.aps.org/doi/10.1103/PhysRevA.111.042809}
}

@article{Frighetto2025,
  title = {Improved positron-molecule scattering calculations with the Schwinger multichannel method},
  author = {Frighetto, Francisco F. and Barbosa, Alessandra Souza and Sanchez, Sergio d'A.},
  journal = {Phys. Rev. A},
  volume = {112},
  issue = {4},
  pages = {042803},
  numpages = {12},
  year = {2025},
  month = {10},
  publisher = {American Physical Society},
  doi = {10.1103/bh8d-drw5},
  url = {https://link.aps.org/doi/10.1103/bh8d-drw5}
}

@misc{Gregg2025Gamma,
      title={Many-body theory and Gaussian-basis implementation of positron annihilation $\gamma$-ray spectra on polyatomic molecules}, 
      author={S. K. Gregg and J. P. Cassidy and A. R. Swann and J. Hofierka and B. Cunningham and D. G. Green},
      year={2025},
      eprint={2502.12364},
      archivePrefix={arXiv},
      primaryClass={physics.atom-ph},
      url={https://arxiv.org/abs/2502.12364},
      note={arXiv:2502.12364 [physics.atom-ph]}
}

@misc{Gregg2025,
      title={Many-body theory calculations of positron binding to parabenzoquinone}, 
      author={S. K. Gregg and J. Hofierka and B. Cunningham and D. G. Green},
      year={2025},
      eprint={2502.10327},
      archivePrefix={arXiv},
      primaryClass={physics.chem-ph},
      url={https://arxiv.org/abs/2502.10327},
      note={arXiv:2502.10327 [physics.chem-ph]}
}

@book{mbtexposed,
author = {Dickhoff, Willem H and Van Neck, Dimitri},
title = {Many-Body Theory Exposed!},
publisher = {World Scientific},
year = {2025},
doi = {10.1142/14178},
address = {},
edition   = {3rd Ed.},
URL = {https://www.worldscientific.com/doi/abs/10.1142/14178},
eprint = {https://www.worldscientific.com/doi/pdf/10.1142/14178}
}

@misc{Rosario2026,
      title={Coupled cluster theory for positron binding in anions and polyatomic molecules}, 
      author={Rosario R. Riso and Jan Haakon M. Trabski and Federico Rossi and Dermot Green and Henrik Koch},
      year={2026},
      eprint={2603.19948},
      archivePrefix={arXiv},
      primaryClass={physics.chem-ph},
      url={https://arxiv.org/abs/2603.19948},
      note={arXiv:2603.19948 [physics.chem-ph]}
}

@article{VMD1996,
    author={Humphrey, W. and Dalke, A. and Schulten, K.},
    title={{VMD - Visual Molecular Dynamics}},
    journal={J. Molec. Graphics},
    year={1996},
    volume={14},
    pages={33},
    url={http://www.ks.uiuc.edu/Research/vmd/}}

@MISC{NWChem,
title = {https://www.nwchem-sw.org/},
url = {https://www.nwchem-sw.org/},
year={2024}
}

@misc{kelvin2,
title = {https://www.ni-hpc.ac.uk/},
url = {https://www.ni-hpc.ac.uk/},
year = {2025}}

@misc{basis_set_exchange,
year = {2024},
howpublished = {https://www.basissetexchange.org/}
}




  
  

\end{document}